\documentclass[]{article}

\usepackage{color}
\usepackage{graphicx}
\usepackage{amsmath}
\usepackage{hyperref}
\usepackage[normalem]{ulem}
\usepackage{etoolbox}
\usepackage{tabularx}
\usepackage[margin=1.0in]{geometry}
\usepackage{multicol}
\usepackage{url}
\usepackage[titletoc]{appendix}
\usepackage{longtable}
\def\Msol{$\rm M_\odot$\ }

\definecolor{spring}{rgb}{0.7,0.9,0.7}
\definecolor{brick}{rgb}{0.7,0.2,0.1}
\definecolor{redHL}{rgb}{1.0,0.5,0.5}

\begin{document}

\title{Next Generation Observatories \\
\large Report from the Dawn VI workshop; October 5-7 2021\\
}
\author{Dawn VI SOC and Presenters}

%%%%%%%%%%%%%%%%%%%%%%%%%%%%%%%%%%%%%%%%
% abstract
%\begin{abstract}
%\input{abstract}
%\end{abstract}

\maketitle

\tableofcontents
\newpage

%%%%%%%%%%%%%%%%%%%%%%%%%%%%%%%%%%%%%%%%
% Preface
\section{Preface}
The workshop \textit{Dawn VI: Next Generation Observatories} took 
place online over three days, 5-7 October, 2021. More than 200 physicists and astronomers attended to contribute to, and learn from, a discussion of next-generation ground-based gravitational-wave detectors.  The Dawn meetings\footnote{The previous Dawn meeting presentations and reports can be found at \url{https://gwic.ligo.org/conferences.html}} were suggested in 2015 by the National Science Foundation (NSF) to provide a forum for the community interested in gravitational wave observational and instrument science, and have been vital in establishing a consensus view of the path forward for US work in the field, and more broadly for work on next-generation science. The Dawn VI focus was the US plans for a next-generation observatory, Cosmic Explorer.

The LIGO and Virgo Collaborations had completed the Third Observing Run (O3) in March 2021 with public alerts indicating numerous black hole binaries, several neutron star binaries, and two neutron-star -- black-hole binaries. Within weeks of the Dawn VI workshop,  the Catalog GWTC-3\footnote{\url{https://arxiv.org/abs/2111.03606}} was published, detailing the detection of gravitational waves from a total of 90 astrophysical sources. The inferences from the multi-messenger observation of GW170817 have continued to grow. KAGRA has joined with Virgo and NSF's LIGO detectors to share observing data. Incremental improvements in the LIGO, named `A+'\footnote{\url{https://arxiv.org/abs/1410.5882}, \url{https://arxiv.org/abs/1304.0670}} and the Virgo `AdV+' project\footnote{Proc. SPIE 11445, 1144511 (13 December 2020); \url{https://doi.org/10.1117/12.2565418}} instruments in the current 3\,km and 4\,km  observatories are well underway, with the start of the next observing run O4 of LIGO, Virgo, and KAGRA planned for late 2022. LIGO and Virgo are starting to consider possible further improvements in the currently operating observatories, but there are practical limits to sensitivity given the 4 and 3\,km length of the arms.

Excitement from scientists and the general public 
 world-wide has further fueled the exploration of ideas and plans for the  next generation of terrestrial gravitational-wave observatories. The 2020 Astro Decadal Survey\footnote{\url{https://www.nap.edu/catalog/26141/pathways-to-discovery-in-astronomy-and-astrophysics-for-the-2020s}} spoke highly of the scientific value of improved instruments, and the potential of Multi-Messenger Astrophysics (MMA) including gravitational waves was recognized as a priority by a very wide range of scientists. Both Einstein Telescope, a European concept, and Cosmic Explorer, a US concept, have made significant progress toward realization in the last few years.
 
During the three-session workshop, sessions and discussions explored the context and the concepts around these next-generation observatories, organized around the following topics:
\begin{itemize}
\parskip -3pt
\item Observational Science opportunities (Section~\ref{sec:survey})
\item Detector and Observatory designs (Section~\ref{sec:design})
\item Realizing designs and evolving practice (Section~\ref{sec:projects})
\end{itemize}

This report represents the views of those participants at the workshop and other interested colleagues who reviewed the document and endorsed it.

%\footnote{These individuals are listed in Appendix~\ref{appendix}.}. \\
\bigskip
\noindent\textit{
David Shoemaker, on behalf of the Scientific Organizing Committee:}

\begin{table}[!ht]
\begin{tabular}{l l}

Stefan W.	Ballmer	&	Syracuse University	\\
Matteo Barsuglia & CNRS  - APC, Paris \\
Josh Frieman & Fermilab \\
Mansi Kasliwal & Caltech \\
Jackie Hewitt & MIT \\
Chuck Horowitz & Indiana University \\
Brian Lantz & Stanford \\
Albert Lazzarini & Caltech \\
David Shoemaker, Chair & MIT \\

\end{tabular}
\end{table}

\newpage

\noindent Dawn IV report section presenters and round-table organizers: 

\begin{table}[!ht]
\begin{tabular}{l l}

Stefan W.	Ballmer	&	Syracuse University	\\
Emanuele	Berti	&	Johns Hopkins University	\\
Patrick Brady & University of Wisconsin-Milwaukee\\
Duncan Brown &	Syracuse University	\\
Poonam Chandra &	National Centre for Radio Astrophysics, TIFR, India	\\
Matt Evans & MIT \\
Ke Fang & University of Wisconsin-Milwaukee\\
Wen-fai Fong & CIERA \& Northwestern University \\
Peter Fritschel & MIT \\
Harald L\"uck &	Leibniz Universität Hannover	\\
Hiranya Peiris &	University College London / Stockholm University	\\
Michele Punturo & INFN Perugia \\
Dave Reitze & Caltech \\
Gary H. Sanders & Simons Observatory, Center for Astrophysics and Space Sciences, UC San Diego  \\
Bangalore Sathyaprakash &	Penn State and Cardiff University	\\
Bram J. J. Slagmolen    &   Australian National University and OzGrav \\
Joshua Smith &	California State University, Fullerton	\\
Andrew Steiner & University of Tennessee, Knoxville and Oak Ridge National Laboratory \\
Eleonora Troja & University of Rome Tor Vergata \\
V. Ashley Villar & Penn State \\
Rainer Weiss & MIT \\

\end{tabular}
\end{table}

\begin{multicols}{3}
\begin{description}
\parskip=-3pt
%    \item Emanuele Berti
%    \item Duncan Brown    
%    \item Poonam Chandra
%    \item Eleonora Troja
%    \item Hiranya Peiris
%    \item Andrew Steiner
%    \item Ke Fang
%    \item Wen-fai Fong
%    \item Bangalore Sathyaprakash
%    \item Peter Fritschel
%    \item Harald L\"uck
%    \item Michele Punturo
%    \item Stefan Ballmer
%    \item Joshua Smith
%    \item Dave Reitze
%    \item Patrick Brady
%    \item Gary Sanders

%    \item Matthew Evans
    \item   
    \item
    \item
    \item
\end{description}
\end{multicols}

\newpage

%%%%%%%%%%%%%%%%%%%%%%%%%%%%%%%%%%%%%%%%
% Executive summary
\section{Executive summary}

\textbf{The workshop \textit{Dawn VI: Next Generation Observatories} took place online over three days, 5-7 October, 2021. More than 200 physicists and astronomers attended to contribute to, and learn from, a discussion of next-generation ground-based gravitational-wave detectors.} 

The program was centered on the next generation of ground-based gravitational-wave observatories and their synergy with the greater landscape of scientific observatories of the 2030s.  Cosmic Explorer (CE), a concept developed with US National Science Foundation support, was a particular focus; Einstein Telescope (ET), the European next generation concept, is an important complement and partner in forming a network.  

The first day, the organizers invited speakers from a range of astrophysics disciplines, with the objective of understanding how gravitational wave observations can complement and enable progress in those non-GW disciplines. Presentations were given on the synergy of GWs with radio astronomy, gamma/x-ray observations, nuclear physics, visible and near IR observations, and particle physics. The possibilities for cosmology, and for multi-band and isolated gravitational-wave observations, were also discussed, along with the transition from the current state of the detectors (in preparation for a late-2022 start for the O4 observing run) to the time when the next-generation observatories of CE and ET can commence operations. \textbf{There was a clear sense of opportunity and potential for combining GW and non-GW messengers to realize otherwise inaccessible science. The example of GW170817, the binary neutron-star merger whose excellent sky localization, provided by LIGO and Virgo, allowed it to be studied in GWs, gamma rays, and multiple additional photon wavelengths, was repeatedly invoked to stress the importance of a 3 (or more) GW detector network. }

The second day, the instruments were the focus. The work in the LIGO and Virgo collaborations to explore upgrades between the end of the current instrument plan (end of O5, $\sim$2028) to the start of the CE/ET epoch ($\sim2035$) was discussed. The ET and CE Projects were described, with CE as the focus of the meeting in more detail. \textbf{The design for CE stresses low initial technical risk, with an instrument design closely based on NSF's  Advanced LIGO detector, and increasing the length of the interferometer arms to increase sensitivity}; it trades away technical risk for a larger infrastructure, and value engineering of the vacuum system is a strong focus of the team. The perspectives of organizational entities that played key roles in the success of Advanced LIGO were provided: LIGO Laboratory, LIGO Scientific Collaboration, and the NSF. 

The final day featured a discussion of very large science projects, of the scale of CE (\$1 - \$2 billion USD). Experienced scientists who have managed `megaprojects' for NASA, DOE, and NSF offered insights for CE. Assembling an approach to funding may call on multiple funding agencies, potentially on private foundations, and international collaboration will be an early challenge. \textbf{Finding an organizational structure which can credibly manage the Project must be an early goal of the CE team. Site selection is complex, time consuming, and key. Establishing a real integration of the CE team with the Indigenous Peoples on whose land the sites will lie is imperative.} Given the duration of the Project, a pipeline of early career scientists is crucial; related is the importance of forming a core team from a broad range of institutions. \textbf{The Cosmic Explorer Project must create a diverse, inclusive, and equitable environment to both engender and profit from a full spectrum of participation.}The bottom line is that a project of this scale will be rife with challenges, but the goal is clearly of sufficient value to merit the effort to make it happen.

A round table on data processing and access models for the CE epoch followed. CE plans to keep within the scope of the Project the effort to characterize the detectors, to calibrate the data, to `clean' the data of defects, and to issue low-latency alerts to the scientific public. The alerts will carry all available information to aid in prioritization by non-GW observers. CE will require greater automation of data conditioning than is currently taking place with the second generation detectors: however both LIGO and Virgo will be focusing on this during the coming years, so it is expected that they will provide a legacy upon which to build by the time of CE observations. \textbf{The full CE data stream is planned to be released as soon as it is in a suitable form for analysis, without a proprietary period.} This is a different approach than the current LIGO-Virgo-KAGRA practice, where the data are also analyzed deeply by the Collaborations before release, but feels to the CE team to be forward-looking and the best assurance of the broadest possible participation in Cosmic Explorer's science. To properly analyze the data, progress will be needed in waveform generation and parameter estimation methods. The computing \textit{per se} does not look excessively difficult, with few overlapping signals anticipated and with anticipated progress in computing speed in the coming 15 years. 

The concluding summary of the meeting expressed the sentiment that the observational science accessible to CE and ET, also in combination with data from other non-GW observatories, will stimulate a very broad community of analysts and yield insights which are exciting given the access to GWs from the entire universe. The need, and desire, for closer collaboration between ET and CE was expressed; a three-detector network is optimal for delivering much of the science. 
\bigskip

\textbf{The science opportunities afforded by CE and ET are broad and compelling, impacting a wide range of disciplines in physics and high energy astrophysics. There was a consensus that CE is a concept that can deliver the promised science. A strong endorsement of Cosmic Explorer, as described in the CE Horizon Study, is a primary outcome of DAWN VI.}

\subsection{Recommendations}

\textbf{Observational Science}

\begin{itemize}

	\item Coordination between the current generation of detectors is crucial. Once KAGRA and LIGO-India are at a sensitivity level comparable to LIGO and Virgo, phased upgrades while maintaining a 3-detector network will be possible. The planning should have input from the non-GW observing community and be take into account limited-lifetime space missions.
	\item A network of next-generation detectors is important to realizing the science goals of the community since sky localization is crucial for MMA. This point came up throughout the meeting, and the strong consensus recommendation is that three (or more) detectors, closely coordinated in observing and upgrading, must be the goal. 
	\item Discussion is needed between the U.S. and E.U. projects on the path forward for data access in the epoch of CE and ET. Realizing the network will require a common policy. 
	\item A number of topics requiring one-time or sustained communication between GW and non-GW observers led to recommendations for meetings: 
	\begin{itemize}
    	\item Explicit coordination of GW observation with (in particular) space missions worldwide, due to their limited lifetime, is needed to ensure that MMA can be best realized.
    	\item A regular interaction is needed between the GW and EM/particle/nuclear domains to ensure that scientific and strategic insights which can impact any of the domains are communicated clearly and in a timely way. A small group that has visibility and credibility with all communities could be effective.
    	\item The scientific benefit of precise localization by a GW network, and the need for low latency vs. high latency with improved precision, should be discussed in a Town Hall to develop a common understanding and consensus.
    	\item A discussion is needed on how (or if) additional low latency GW information can be used to support triage by EM/particle observers to manage the higher event rates in O4 and beyond. 
    	\item The long-standing tension between the duty-cycle of observing vs. improving the sensitivity with its concomitant downtime continues, covering the upgrade arguments (net event count and signal-to-noise ratio (SNR)) vs. the value of extended observation at a fixed sensitivity. Post-O5 upgrades should be included in the scope of discussion.
	\end{itemize}
\end{itemize}

\textbf{Instrument and Project Recommendations}

\begin{itemize}
    \item We recommend the establishment of a small group charged to explore and launch efforts to initiate specific limited coordinated activities between CE and ET such as model code and parameters, and vacuum equipment value engineering.
    \item To foster the engagement of the LIGO Lab in CE, discussions should begin soon among the CE team, the LIGO Laboratory management team, and the NSF.
    \item It is urgent that the CE project design phase take place within an organizational structure developed as a precursor that can evolve to become the eventual, robust facility organization.
    \item The CE design stage should move early to a baseline configuration and fully develop that design rather than continuing the trade studies of potential design option unless consideration of options is found to be required to manage uncertainty and risk.
    \item A group should be identified or established to coordinate GW and EM/particle strategic planning, to ensure the best use of time-constrained facilities. 
    \item CE should establish now a Project Advisory Committee with membership including expertise in similar-scale projects.
    \item GW facility leadership (both present and planned) should develop a long-term ``mission statement” relevant to lifetimes of space missions dedicated to GW follow-up. This requires a decade-scale horizon for the planning of instrument operation, to match the time scale for space missions.
    \item  The observational science value to having a network node in the southern hemisphere is significant . The community should continue to explore means to realize a next-generation observatory there.
    
\end{itemize}

\newpage

%%%%%%%%%%%%%%%%%%%%%%%%%%%%%%%%%%%%%%%%
% Survey of progress 
\section{Survey of progress}\label{sec:survey}
\subsection{Progress in the field between Dawn IV and Dawn VI} 

Dawn IV was held in August 2018\footnote{\url{https://indico.nikhef.nl/event/1174/};\url{https://gwic-documents.s3.us-west-2.amazonaws.com/dawn/Dawn-IV-report.pdf}}, and Dawn V\footnote{\url{https://indico.ego-gw.it/event/20/}} in May 2019 (this Dawn VI report also covers Dawn V). Following the O2 Observing run which ended in August 2017, NSF's LIGO and the European Virgo detectors were brought to an improved sensitivity, through the use of squeezed states of light (improving the high-frequency shot-noise limited sensitivity for a given laser power) and many technical advances. The joint Virgo-LIGO O3 Observing run, April 2019 to March 2020, yielded some 79 new events. Very massive black holes, compact objects straddling the mass range between neutron stars and black holes, and mixed black-hole -- neutron-star mergers were among the events. The catalog GWTC-3\footnote{\url{https://arxiv.org/abs/2111.03606}} provides an overview and pointers to further papers. KAGRA came on line with a limited sensitivity toward the end of this period and jointly observed with GEO-600. 

In parallel, improvements to the sensitivity of LIGO and Virgo were planned, funded, engineered, and are being implemented at the end of 2021. The LIGO `A+' and Virgo `AdV+' upgrades add frequency-dependent squeezing (to allow broadband quantum noise improvement) as well as a range of other changes (higher laser power, additional baffling, replacement of some mirrors, etc.) intended to improve the reach of the instruments. The next run, O4, is expected to start in late 2022 once the upgraded detectors have been initially commissioned. 

The GWIC 3G Committee reports, which were in draft form for Dawn IV, have been refined, and are available on the GWIC web pages\footnote{\url{https://gwic.ligo.org/3Gsubcomm/documents.html}} or the arXiv\footnote{\url{https://arxiv.org/abs/2111.06991, https://arxiv.org/abs/2111.06990, https://arxiv.org/abs/2111.06989},\\
\url{https://arxiv.org/abs/2111.06988, https://arxiv.org/abs/2111.06987, https://arxiv.org/abs/2111.06986}}. These reports aided greatly in the preparation of the proposal for the European Einstein Telescope (ET) proposal to the ESFRI process, as well as for the preparation of the Cosmic Explorer Horizon Study (CEHS)\footnote{ibid}. ET was placed on the ESFRI (European Strategic Forum for Research Infrastructures) roadmap; the CEHS was completed and submitted to the NSF.

\textbf{The US Decadal Survey on Astronomy and Astrophysics 2020 (Astro2020 Decadal\footnote{\url{https://www.nap.edu/catalog/26141/pathways-to-discovery-in-astronomy-and-astrophysics-for-the-2020s}}) Study took place in this interval, and produced its draft report shortly after the Dawn VI meeting. It was informed by many white papers from the astronomical community including the gravitational-wave field -- both for instrumentation and for observational science. The report speaks highly of the value of gravitational-waves as an astronomical messenger both stand-alone and in multi-messenger astrophysics. The report ``strongly endorses investments in technology development for advanced gravitational wave interferometers, both to upgrade NSF’s Laser Interferometer Gravitational-Wave Observatory (LIGO), and to prepare for the next large facility.'' Further, ``The survey committee strongly
endorses gravitational wave observations as central to many crucial science objectives''.} 

\subsection{Status of Recommendations from Dawn IV}

\begin{itemize}
\item \textit{GWIC should found an international Umbrella Organization by the Dawn V meeting in Spring 2019 to coordinate international research and development for 3G and detector upgrade plans.} This topic was discussed in some depth at Dawn V, and the conclusion was that the two primary next generation observatory projects (CE and ET) were in a sufficiently different state of maturity that it might not be productive to couple the projects closely at that time. The current observatories and collaborations are starting to share more computing and communications infrastructure through IGWN (International Gravitational-Wave Network), and informal technical exchanges are growing between ET and CE. 

\item \textit{The ground-based GW community should prepare to respond to calls for input to roadmaps. Specifically for the US Astro2020 Decadal Survey, we should respond through the submission of i) a coordinated set of science white papers, and if required by the Astro2020 charter ii) 
roadmaps for development of mid-scale technologies and programs to enable 2.5 and 3G detectors. 
A proposal should also be included in the 2021 European Strategic Forum for Research Infrastructures (ESFRI) roadmap.} White Papers were indeed prepared for the Astro2020 Decadal survey, both on the general topic of gravitaional waves as a meessenger, and more specifically to inform the community about CE; the Report (released in early November 2021) indicates that those White Papers were quite influential in the preparation of the report. ET did propose to the ESFRI program and was adopted in Summer 2021 to the ESFRI Roadmap. 

\item \textit{A Dawn V meeting should be held when the GWIC-led 3G subcommittee report is expected to be released to the  community. The focus should be the 3G subcommittee report and its use for informing international funding agencies, including the the US Astro2020 Decadal Survey and the 2021 ESFRI roadmap in Europe.} Dawn V was held 26-26 May 2019\footnote{\url{https://indico.ego-gw.it/event/20/}}. The 3G reports were in good draft condition at the time of the meeting, and the approach of the community for the US Astro2020 Decadal Survey and the 2021 ESFRI roadmap in Europe was discussed. A range of papers and proposals were created using the GWIC 3G reports, and this was directly helpful in the success of ESFRI adoption of ET and in the preparation of the CE Horizon Study with recognition by the Decadal Committee. 

\item \textit{Exploring the astrophysical science gain of a third 3G facility and placement in the southern hemisphere should be a top priority for GWIC.} The GWIC 3G Science Book carried out an analysis of the advantages of a southern-hemisphere detector, and the CE Horizon Study included evaluation of an Australian site. An Australian study for an observatory with a focus on the end-phase of neutron-star mergers (NEMO\footnote{\url{https://arxiv.org/abs/2007.03128}}) has made significant progress. \textbf{Further work on the value of and means to realize a next-generation observatory in the southern hemisphere is recommended.}

\item \textit{We recommend that the 3G science case evaluate the science contribution from below 10~Hz versus above 10~Hz for each item, to help inform detector requirements, and for 500~Hz to 4~kHz, potentially from 2G and 2.5G detectors with shorter baselines of 3-4~km.} This study was pursued first for the GWIC 3G documents, and then in more detail for the CE Horizon Study. The outcomes were applied to the trades for the CE baseline design, and led to the conclusion that CE should consist of two widely separated observatories of 40km and 20km length to span the anticipated sources. The specific value as a function of lower frequency cutoff was explored in the ET ESFRI submission materials. The field has a clearer vision of the observational science as a function of frequency span. 

\item \textit{The GW community should address the development of software and computing hardware in parallel with the instrumentation and science development.} The preparation of both the ET ESFRI proposal and the CE Horizon Study provided a motivation and opportunity to work on the requirements for next-generation observatory observational science needs for computation and for refinement of the waveform models required to exploit the very high signal-to-noise ratio one can expect for some events. In addition, Astro2020 Decadal White Papers brought home the need for continued work (and funding!) for this domain. The continuum of improvement in the current Virgo, LIGO, and KAGRA detectors is also stimulating progress there. 

\item \textit{The 3G community should adopt the following common strategies: }
	\begin {itemize}
	\item \textit{Establishment of a common research and development program within the U.S. and Europe to facilitate the exchange of information and optimize the global expenditure of efforts.} This has not happened. There has been some exchange of thinking at meetings, but between the tight focus on the ET ESFRI proposal and CE Horizon Study, and past practice, this remains an open action item.
	\item \textit{As part of a broader global research and development effort, investment in more global resources devoted to characterization of coatings at cryogenic temperatures, such as a dedicated, internationally resourced coatings center.} No central research center has been developed. However, there has been greater collaborative effort between the European and US efforts, spanning the applications from current to CE and ET. Yet closer integration of the efforts would be advantageous. 
	\item \textit{Global coordination of prototype engineering and scaled tests for 3G detectors, including beam tube construction, vacuum technologies, and excavation and construction methods.} Some progress has been made here. The ET Pathfinder Laboratory, in Maastricht, Netherlands, was funded and has just been inaugurated in late 2021. It is intended to provide an environment that can allow full-scale testing of components of next-generation observatory detectors, and explicitly wishes to serve both ET and CE. Joint discussions on approaches to lower-cost vacuum systems have been undertaken, with one in-person meeting including CERN and US vacuum experts (the next joint meeting is tentatively scheduled to be hosted by CERN in the fall of 2022), and some initial research on surface treatments underway. There has been little global discussion on excavation, tunneling, and general civil construction to date, and there are fewer evident commonalities with the present conceptual designs. More effort in all these areas would be fruitful. 
	\item \textit{Development of a long term plan that balances observing with installation and commissioning breaks to make use of current generation facilities as a testbed for 3G technologies.} Both Virgo and LIGO have launched studies of the use of the present 3- and 4-km infrastructures to help explore next-generation detector designs. There is good communication but no formal coordination between the technical studies, and this should be encouraged. 
	\end{itemize}
\item \textit{The 3G community should continue to explore paths from the current organization of largely independent projects toward a global unified endeavor, with the objective to optimize the use of financial and human resources, and to maximize the science from a 3G network.} Dawn V attempted to stimulate discussion for the form that a common organization could take, but at that time there was little enthusiasm to adopt any overarching structure. The need for each of ET and CE to focus on important deadlines (ESFRI, CE Horizon Study) played a role, combined with a continuing desire for the lack of constraints in the current approach. Dawn VI chose to focus on the US Cosmic Explorer concept, and did not confront this question. In parallel, there is some effort between the current LIGO Scientific Collaboration, Virgo Collaboration, and KAGRA Collaboration to bring some of the common computing effort under an informal organization of `IGWN' (International Gravitational-Wave Network). This may be a kernel which can grow to cover a wider scope, but in any event serves to show the advantages of shared resources and common management in a limited domain. 
\item \textit{The GW, EM, and neutrino communities should coordinate to identify key joint science targets for multi-messenger studies.} The O3 Observing run,  April 2019 to March 2020, resulted in many triggers to the EM and neutrino communities. While a number of events were followed up, no new `MMA' (multi-messenger astrophysics involving gravitational waves) events were identified. Through followups, joint papers involving all partners, and White Papers for the Astro2020 Decadal, more planning and proposing along the lines of coordination across the MMA field has taken place. 

\item \textit{Lastly, we recognize that the GW community is currently dependent on a single optic coating facility (LMA) for detectors present and future, and a single source for any critical component is viewed as a potential risk to the whole community. In addition to the recommendations for enabling a next-generation global detector network above, we also highly recommend the re-establishment of the coating capabilities in hardware and staffing once offered by CSIRO in Australia to mitigate this present risk and as a significant contributor to broader ongoing research efforts.} The Australian National University has funded the creation of a laboratory for coating and the CSIRO equipment has been installed there and is undergoing commissioning at the end of 2021. In addition, LMA has established an oversight committee including a US member and has worked in close collaboration with both US and European researchers in the evolution of coatings targeted for detectors in the current observatories. 
\end{itemize}
\pagebreak

\subsection{Gravitational-Wave International Committee Perspective}

The Gravitational Wave International Committee (GWIC) was formed in 1997 to facilitate international collaboration and cooperation in the construction, operation and use of the major gravitational wave detection facilities world-wide. It is associated with the International Union of Pure and Applied Physics as its Working Group WG.11. Through this association, GWIC is connected with the International Society on General Relativity and Gravitation (IUPAP's Affiliated Commission AC.2), its Commission C19 (Astrophysics), and another Working Group, the AstroParticle Physics International Committee (APPIC). 

GWIC meets yearly to assess the state of the field, and to launch initiatives intended to help facilitate the growth of the field. The `3G' reports which helped ET and CE are a good example; another is a recent publication in Nature Reviews Physics of ``Gravitational-wave physics and astronomy in the 2020s and 2030s\footnote{\url{https://www.nature.com/articles/s42254-021-00303-8}}'', a roadmap for the future of the field. 

Here we very briefly summarize the activities in the field to provide context for the Dawn VI notes and recommendations.

\subsubsection{Ground-based Gravitational-wave Detectors}\label{sec:gwic-ground}

Current ground-based gravitational-wave detectors include KAGRA, Virgo, GEO-600, and LIGO. All of these instruments use variations on the idea of a Michelson interferometer, illuminated with a laser, and with test masses well isolated from external forces. A quadrupolar gravitational wave induces a differential phase shift in the light propagating from the beamsplitter to the end test mass which serves as a mirror and upon recombining at the beamsplitter acts as a transducer, converting gravitational-wave strain into changes in light intensity which then can be converted into an electrical signal. As the gravitational wave is a strain in space, $h=\Delta L/L$, longer arms $L$ lead to larger signals for gravitational wavelengths shorter than the arm lengths. KAGRA and Virgo have 3\,km-long arms, and LIGO 4\,km. Isolating the test masses from external forces leads to a range of approaches. KAGRA is underground, reducing the effect of surface seismic excitation. Virgo and KAGRA use many pendulums in series to act as a mechanical low-pass filter, attenuating seismic motion. LIGO uses sensors and high-gain servo systems in series with pendulum stages to attenuate mechanically-coupled seismic motion. Thermal motion of the test mass can also mask gravitational waves, and low-loss materials for the test mass and optical coatings are used to concentrate the motion in a limited frequency range. KAGRA also plans to reduce the temperature to tens of Kelvin to reduce the thermally-driven motion.

Larger laser powers improve the sensitivity, by improving the Poisson counting statistics. 100W-class lasers are used, and recently prepared states of light -- squeezed light -- have been introduced to manage the quantum noise. The present optical systems, while based on a Michelson interferometer, are quite complex to make the best use of the light and to match the constraints of the observatory infrastructures. The light must travel in an ultra-high vacuum to avoid scintillation in the path, and much of the cost of the observatories is in the providing the vacuum system, its support, and its protection. 

Thus far Virgo and LIGO have achieved a sufficient sensitivity to detect gravitational waves, and some 90 candidates have been observed to date. Programs to further refine the sensitivity are underway, and the next observing run is anticipated for late 2022. The observatory infrastructures, originally built in the 1990's, can continue to be used for some time yet if properly maintained, and there are presently concepts for making incremental increases in the sensitivity of the detectors. 

However, making a significant step forward in sensitivity will require moving to larger arm lengths $L$ to make larger phase shifts in the light for a given gravitational-wave strain $h$. The Dawn VI meeting focused on the US plans for such a revised observatory infrastructure, Cosmic Explorer, where 20km and 40km long arms are proposed. 

The target frequency range for ground-based detectors is from several Hz (limited by the Newtonian background) to several kHz (limited by the current and likely future practical constraints of laser interferometry). 
 
\subsubsection{Space-based Gravitational-wave Detectors}\label{sec:gwic-space}

Space-based interferometer gravitational-wave detectors have a significant attraction in avoiding the stray forces that come from seismic motion, and the Newtonian background due to time-varying density changes near the test masses. In addition, the vacuum path for the light is `free' once the complexity and expense of putting the detector in space is made. 

The LISA (Laser Interferometer Space Antenna\footnote{\url{https://www.cosmos.esa.int/web/lisa/lisa-documents}}) is the most advanced of the concepts for this approach. It is an ESA-led mission, with NASA as a junior partner, to place three satellites in a triangular configuration, and to sense the motion of shielded test masses using the timing of laser beams passed between pairs of satellites. The detector targets inspirals of $10^6$-$10^9$ solar mass binaries, with some sensitivity for less massive systems. The planned armlength is $2.5\times 10^9$ meters, which is comparable to the gravitational-wave wavelengths in the mid-band of sensitivity. At low frequencies (around $10^{-7}$\,Hz) it is limited by residual forces on the imperfectly-shielded test mass, and at high frequencies (around $0.1$\,Hz) by the smaller net path-length change when multiple wavelengths of the gravitational-wave fit in the arm length. It is planned for a mid-2030's launch.

In Japan, the gravitational wave community agrees that DECIGO will be the next main project after KAGRA. DECi-hertz Interferometer Gravitational-wave Observatory (DECIGO) is a future Japanese space-based gravitational-wave antenna that will target the detection of an early-universe stochastic gravitational wave signature imprinted by inflation. 

\subsubsection{Pulsar Timing Array Gravitational-wave Detectors}\label{sec:gwic-PTA}

A Pulsar Timing Array (PTA\footnote{Hellings, R. W. \& Downs, G. S. Upper limits on the isotropic gravitational radiation background from
pulsar timing analysis. Astrophys. J. Lett. 265, L39–L42 (1983).}) measures variations in the radio frequency pulse arrival times at the Earth from an array of millisecond pulsars.  Pulsars are highly-magnetized neutron stars, spinning on an axis due to angular momentum acquired from the original pre-collapse star. Many pulsars are observed to be extraordinarily accurate clocks. As a gravitational-wave passes between the Earth and a pulsar, the time-of-flight of radio waves is modulated, and this then can be interpreted as an observation of gravitational waves. In practice, many pulsars are observed, and one seeks to find the quadrupolar signature of GWs in the spatial distribution of phase shifts. The distances to pulsars (the closest is some 400 light years distant) makes this approach best suited to signals around $10^{-9}$ Hz, and so mergers of the largest BH at centers of galaxies. Data have been accumulated with this goal in mind over several decades, and first indications of a signal may be appearing\footnote{See, for example, \url{https://arxiv.org/abs/2005.06495}}.

\subsection{GW and MMA Community Planning Activities}

There is a significant level of activity in the broader community to look for synergies and to plan for future facilities which supports both the GW detectors and the multi-messenger astrophysics that can be achieved. We list here some of these activities:

\begin{itemize}
    \item Snowmass\footnote{\url{https://snowmass21.org/}}: The Snowmass/P5 process is a two-step process used by the particle physics community and its U.S. funding agencies, Department of Energy (DOE) and National Science Foundation (NSF) to formulate the strategic plan for U.S. particle physics. Snowmass is the first step in the planning process. It is a community-driven study of the scientific opportunities organised by a number of the American Physical Society’s Divisions: DPF, DPB, DAP, DNP, and DGRAV. The 2022 Snowmass Community Summer Study will be held at the University of Washington from 17-22  July  2022.
    \item Aspen\footnote{\url{https://www.aspenphys.org/physicists/summer/program/currentworkshops.html}}: The Aspen Center for Physics is hosting a meeting 5 June-3 July, 2022, on `Fundamental Physics and Astrophysics with the Next Generation of Gravitational-Wave Detectors'. The chief goal of this Aspen workshop is a cross-disciplinary extensive debate and careful planning to prepare for the data from the next generation GW observatories. 
    \item GWADW\footnote{Queries to \href{mailto:gwadw2022@gw.phys.titech.ac.jp}{\texttt{gwadw2022@gw.phys.titech.ac.jp}}}: The Gravitational-Wave Advanced Detector Workshop series is an annual meeting, typically held in a relatively isolated locale, where a true workshop approach is taken to sharing ideas on technologies and instrument science for future detectors. The next meeting is planned for Hokkaido, Japan, from 30 May-6 June 2022. 
    \item GWPAW\footnote{\url{https://gwpaw2021.aei.mpg.de}}: The Gravitational-Wave Physics and Astronomy Workshop is an annual meeting, bringing together GW and other interested domains. The setting is informal, with no parallel sessions and adequate time for discussion. This year the meeting is held December 14-17, 2021. 
    \item PAX\footnote{Summer 2022 meeting will be announced at \url{https://indico.mit.edu/e/PAX2022}; Contact S. Vitale, MIT \href{mailto:svitale@mit.edu}{\texttt{svitale@mit.edu}}}  Physics and Astrophysics at the eXtreme (PAX) is a discussion-based workshop with very strong involvement of participants.  The most recent PAX-VII{\footnote{\url{https://sites.psu.edu/paxvii/}}}, held 23-27 August 2021, was to better understand what the requirements are for the 3G observatory network from different science perspectives, providing useful input to this Dawn VI meeting.  The next PAX is planned to be held at MIT in August, 2022.
\end{itemize}

There is also a broad and deep interest in the MMA community to strive for a diverse, equitable, and inclusive work environment. Given the reality of scientific work, where the boundaries between work \textit{per se} and social life tends to blur, it is all the more important to ensure that social issues are treated as a central element in project and operations planning. To this end, a number of organizations are active in promoting this approach, developing materials and processes to further it, and providing gathering places for all who want to put some time into these issues to contribute. We list here some which are familiar to the MMA domain, and encourage readers to participate in this evolution of the field:

\begin{itemize}
\parskip -3pt
    \item The Multi-messenger Diversity Network\footnote{\url{https://astromdn.github.io}}
    \item APS Inclusion, Diversity, and Equity Alliance\footnote{\url{https://www.aps.org/programs/innovation/fund/idea.cfm}}
    \item IAU subgroup on Management of Diversity and Inclusion in Large International Collaborations\footnote{\url{https://iau-oao.nao.ac.jp/iau-inclusion/management-of-diversity-and-inclusion-in-large-international-collaborations/}}
    \item IGrav\footnote{\url{https://www.igrav.org}} to share best practices among GWIC’s Projects

\end{itemize}
\subsection{Communication with, and between, funding and advisory agencies}

 There is value for GWIC and the working groups in the field to identify and establish communication channels with funding and advisory agencies who currently or may in the future support GW detectors. This provides GWIC a recognized approach to communicate as needed with those agencies on the scientific needs, desires, and constraints from the communities and provides a coherent framework to serve as an advocacy group for the communities.

Groups that GWIC and the community have identified as directly interested in the field include:

\begin{enumerate}
	\item The `Gravitational Waves Agency Correspondents'  
	(GWAC\footnote{\url{https://www.nsf.gov/mps/phy/gwac.jsp}}). GWAC is an informal body of representatives of 
	funding agencies covering ground and space-based 
	gravitational wave projects and science, initially 
	formed in 2015 at the initiative of the US National Science 
	Foundation. Many of the organizations currently supporting the field are represented in GWAC. GWIC has received formal invitations from APPEC and GWAC to present the status and planning activities of the broad gravitational wave community 
that it represents. GWIC representatives attended telecons with 
presentations to GWAC (April 2019, March 2020, April 2021). In those presentations, GWIC communicated the status of 3G 
subcommittee reports and more generally activities in the field. 
GWAC also, at the request of GWIC, provided a very significant review of the `3G Subcommittee' documents. The authors implemented the recommendations in the thoughtful and thorough review, and the finished reports\footnote{\url{https://gwic.ligo.org/3Gsubcomm/documents.html}} reflect the improvements made.
    \item The `Astro-particle Physics European 
	Consortium' (APPEC\footnote{\url{https://www.appec.org}}). This is a consortium of European agencies with responsibility for funding particle astrophysics. APPEC 
	appoints a Scientific Advisory Committee drawn 
	from the European scientific community to maintain a 
	scientific roadmap for particle astrophysics, may choose
	when appropriate to create financial instruments to 
	support strategic areas, and can act via influence to 
	support projects in non-financial actions. The ``Mid-term review of the European Astroparticle Physics Strategy 2017-2026"\footnote{\url{https://www.appec.org/mid-term-review}} contains a chapter on gravitational waves covering current and future ground- and space-born instruments, and the 2022 APPEC Town Meeting is expected to devote significant time to GWs. 
    \item The European Consortium for Astroparticle Theory, EuCAPT\footnote{\url{https://eucapt.org/}}, was established as a centre of excellence hosted at CERN. EuCAPT organizes many events related to GWs such as virtual colloquia or the EuCAPT annual symposium\footnote{\url{https://www.eucapt.org/events}}, and also a white paper (containing GWs) was written\footnote{\url{https://www.eucapt.org/white-paper}}.
    \item The  Joint ECFA-NuPECC-APPEC Activities. (JENAA\footnote{\url{http://www.nupecc.org/jenaa/}}) exists to address common questions in particle, astroparticle and nuclear physics and to better exploit synergies between these fields. It already led to five Expressions of Interest for interdisciplinary cooperation, including one for Gravitational Waves for fundamental physics\footnote{\url{http://www.nupecc.org/jenaa/?display=eois}}
	\item The `Group of Senior Officials' (GSO\footnote{\url{https://www.gsogri.org}}), on Global 
	Research Infrastructures.  The GSO was established and 
	active from 2011 onward to informally explore cooperation 
	opportunities in Global Research Infrastructures (GRIs). 
	Participating countries are represented on the GSO by 
	``government officials and experts in the areas of 
	international research facilities and international relation''. ET has been presented to the GSO at their request.
	\item ESFRI\footnote{\url{https://www.esfri.eu/forum}}, the European Strategy Forum on Research Infrastructures, is a strategic instrument to develop the scientific integration of Europe and to strengthen its international outreach. It supports and benchmarks the quality of the activities of European scientists. ESFRI has the objective of translating political objectives into concrete advice for RI in Europe. ESFRI placed the European Einstein Telescope Project on its roadmap in 2021. 
\end{enumerate}

\newpage
%%%%%%%%%%%%%%%%%%%%%%%%%%%%%%%%%%%%%%%%
% Observational Science
\section{Observational Science}

The organizers invited speakers from a range of astrophysics disciplines, with the objective of understanding how gravitational wave observations can complement and enable progress in those non-GW disciplines. The speakers collected their main points, and comments and questions for the GW field, below.

\subsection{Gravitational Waves}

\textit{Presenter: Emaunele Berti}

The Cosmic Explorer Horizon Study, the inclusion of Einstein Telescope in the 2021 ESFRI roadmap and the strong endorsement that LISA received from the Astro2020 Decadal show that the future of gravitational wave astronomy is bright, and we must think about ways in which these observatories will complement each other.

At PAX-VII\footnote{\url{https://sites.psu.edu/paxvii/}} we tried to understand the requirements for the next-generation observatory \textit{network} from different science perspectives: astrophysical populations of compact objects, black hole binaries from population III stars and beyond, cosmological parameters and backgrounds, waveform and data analysis requirements, tests of gravity and fundamental physics, dark matter candidates, dense matter equation of state and the QCD phase diagram, multimessenger astronomy. This presentation focused on gravitational waves alone, but even in this more limited context, the synergy between multiple GW detectors is crucial.

This is indeed the lesson we learned from the O1, O2 and O3 runs. \textbf{The detections from August 2017 proved that adding the third detector to the network (even if it is somewhat less sensitive than existing detectors) does not just make incremental changes, it can mean a phase transition in the science capabilities of the network}: the rate of detections increased, we were finally able to localize black hole binaries, and the localization of GW170817 marked the beginning of GW multimessenger astronomy.

A comparison of GWTC-2 and GWTC-1 shows that we \textit{learn the most by looking at the corners of the parameter space}: events with large mass ratios like GW190412 and GW190814, the mass-gap event GW190521, or the heaviest binary neutron star GW190425 challenged astrophysical formation scenarios. The most informative events also challenge our understanding of waveform models and they can cause controversial interpretations: What is the nature of the 2.6\Msol compact object? Have we really observed overtones and higher harmonics of the ringdown? Note that “rare" events will be much more common when we get into the CE-ET era of big data. We will mostly learn new physics and astrophysics not by consolidating our understanding of the bulk of the population, but by looking at the unexplored corners of the parameter space.

The large-redshift reach of next-generation ground-based observatories and LISA also means that we will probe \textit{populations that are not accessible now.}
This will allow us to:

\begin{itemize}
\parskip -3pt
    \item probe compact objects that may have primordial or Population III origin; CE and ET will be probing the population \textit{before} the peak of star formation. This gives lots of insight into metallicity, generations of stars (pop III), formation channels, etc. It teaches us about stellar formation and evolution in a unique and powerful way.
    \item do cosmology with standard sirens at higher redshifts, where we could get hints about the resolution of the Hubble tension (if it still holds by the mid-2030s).
\end{itemize}

Finally, the synergy of detectors operating in different frequency bands will probe different possible modifications of general relativity (or different ``new physics”). Multiband sources are best at probing modifications that enter at negative post-Newtonian (PN) orders in the gravitational wave phasing. Massive black holes observed by LISA give better bounds than stellar-origin black holes at negative PN orders, but ground-based observations do slightly better than massive black holes observed by LISA for modifications that enter at positive PN orders. Also, sensitivity improvements in ground-based detectors make a big difference for constraints that appear at negative PN orders. In summary: constraining a variety of possible modifications of GR will also require a synergy between the whole network of Earth and space-based detectors.

Note that much of the science we can do is unlikely to be within reach of other observatories. PTAs or EHT will not detect mergers or probe high-curvature regions. However PTAs, and missions looking specifically for the primordial cosmic gravitational-wave background\footnote{\url{https://cmb-s4.org}} may discover  backgrounds that may not be detectable with ground-based or space-based interferometers.

\textbf{Also to note is that there will be significant progress needed in waveform theory and production, and potentially in analysis coding, to profit from the much greater signal-to-noise ratio and rate expected with the detectors of the future, ground- and space-based. A sustained effort on that front will be well rewarded.}

While the search for B-modes in the CMB is not a topic for this meeting, it is clearly a very exciting avenue for tests of GR and for cosmology as informed by GWs. A direct detection of the time-varying GW strain from e.g., inflationary cosmic gravitational-wave background is inaccessible to the next-generation observatories on the ground and LISA. Space-based detectors targeting this source, perhaps around 0.1-1 Hz, should be the focus of an R\&D effort. In addition, the finite lifetime of space missions invites a series of space-based GW detectors. 

%%%%%%%%%%%%%%%%%%%%%%%%%%%%%%%%%%%%%%%%%
\subsection{Cosmology}
\textit{Presenter: Hiranya Peiris}

In this session was presented an overview of what the landscape of cosmology is likely to be by the end of the Stage IV ``dark energy surveys" in 2030s, and encouraged the focus of Cosmic Explorer to look ahead to the likely breakthrough questions in that era, which move beyond the questions that are the focus of cosmological research today (such as measuring cosmological parameters). There are several avenues with breakthrough potential, including novel signals of phase transitions in the early universe, signatures of the physical nature of dark matter and tests of general relativity that are not feasible using EM data. Finally, one can make a compelling case that \textbf{the most exciting breakthrough potential of Cosmic Explorer for cosmology comes from its discovery potential, particularly through its unprecedented redshift reach.} However, the discovery potential for cosmology needs quantification and clear articulation (within the scientific community and to funding agencies), both to sharpen considerations that affect design decisions and to gather support for funding decisions. While CE has the potential to discover new physics beyond both the standard models of Particle Physics and Cosmology, a major challenge is apparent in quantifying this discovery potential. Beyond-Standard-Model (BSM) signal predictions carry substantial model-dependence, and theoretical uncertainties are significant. Several issues were put forward  for reflection and discussion:

\begin{itemize}
\parskip -3pt
\item How to better define the discovery potential for BSM signals as a function of theoretical uncertainty? e.g., start by defining what broad classes of models can be ruled out?  
\item How to ensure experimental design and data reduction does not inadvertently diminish discovery space while taking into account theoretical and observational advances in the coming decade?
\item How can CE help the theory/modelling community to make forecasts that accurately reflect CE capabilities?
\item How to ensure discovery space viability if coincident EM signals (esp at high z) rare and/or non-existent?
\item Does increased temporal resolution in detector frame for high-z signals enable any novel probes? 
\end{itemize}

The subsequent discussion led off with the question of how the cosmology community currently justifies projects where the dominant breakthrough comes from discovery potential. This is typically done by clearly articulating what broad classes of physical models (as opposed to individual models with tuned parameters) will be ruled out if an experiment which can reach a particular signal-to-noise detects no signal. Then a discussion followed which clarified that the discovery potential at high redshift may be in a regime where EM counterparts are not available or common, and hence a detector network is crucial to preserving discovery potential in this regime.

Once again the importance of a network that can determine the source position in the sky was stressed. 

%%%%%%%%%%%%%%%%%%%%%%%%%%%%%%%%%%%%%%%%

\goodbreak
\subsection{Radio synergy with the Next Generation Gravitational Wave Observatories}

\textit{Presenter: Poonam Chandra}

A true watershed event occurred on 17 Aug 2017, when LIGO and VIRGO registered GWs arising from the merger of two neutron stars in GW170817. This was the first and the only event from which electromagnetic (EM) radiation  was seen from a GW source to date. GWs give information about the mass, spin, and geometric properties like inclination angle, source position and luminosity distance; the EM provides information about the energetics, redshift, host, beaming, environment, ejecta mass and velocity. While the GW 170817 revealed itself in all bands of EM spectrum, it was radio observations which proved quite critical. GW170817 event was instrumental in proving that GW moves with the speed of light, that the merger of neutron stars produce SGRBs and that the heavier elements like gold etc. are synthesized inside during the binary neutron star (BNS) mergers. 

Radio observations are critical in deciphering energetics, environment and evolution. There is a speculation of `prompt' radio emission during the merger; however, it hasn’t been seen yet. If confirmed, it could be a very critical input to fast radio burst (FRB) origin.  On the other hand, the radio ‘afterglow’ emission from compact object mergers involving neutron stars can come via the ejecta interaction with the surrounding medium. During merger a substantial amount of neutron-rich material is ejected. The r-process nucleosynthesis produces a kilonova.  In addition, a jet is launched due to the rapid accretion of ejected matter onto a compact remnant, which will propagate through the merger ejecta medium. This will be the source of non-thermal emission. The interaction of merger ejecta with the surrounding interstellar medium (ISM) produces a long-lasting synchrotron emission observable in multi-wavelength bands; the radio emission is visible for a longer timescale (months and years) and uniquely probes the environment, geometry and the energetics of these systems. An important element in GW170817 was the jet, which transfers a large fraction of its energy into the surrounding ejecta, forming a hot cocoon expanding over a wide angle while traveling at mildly relativistic velocities. As the cocoon propagates into the ISM, it will also produce a radio signal (and other EM signals too).

The clearest evidence of cocoon was seen in GW170817. The rise of the radio light curve at early time was consistent with a model where the radio emission was arising from the subrelativistic cocoon\footnote{2017, Science, 358, 1579, 2018, Nature, 554, 207}. This model could also explain the EM emission in other bands also. The microarcsecond resolution radio observations, i.e., very long baseline interferometry (VLBI) observations, clearly revealed the presence of a jet moving with relativistic speeds\footnote{2018, Nature, 561, 355}. In addition, the modeling of radio observations obtained via worldwide campaign involving multiple telescopes revealing that after the slow rise due to cocoon, the jet did manage to pass through cocoon and revealed itself at late time\footnote{arXiv:2006.02382}. 

As is clear from above that the radio observations in GW170817 provided a plethora of information. While it answered many mysteries, many more questions arose. \textbf{One needs to create a statistically significant sample to understand the diversity of these mergers. For this it is important to focus on next generation GW and EM observatories: The proposed Cosmic Explorer and Einstein Telescope in the GW front and the next generation Very Large Array (ngVLA) and Square Kilometre Array (SKA), are critical towards this venture.} Next-generation GW instruments will be able to detect the BNS mergers up to a redshift of $\ge 1$ and uncover the nature of vast majority of BNS mergers at large distances. Radio observations will provide information on the energy of these mergers and unravel the underlying relation between NS-NS mergers and SGRBs, and provide critical insights into how many such mergers launch a jet. In NS-BH mergers, which may lack the bright optical emission, an optimized radio search strategy is critical to see EM emission from them. 

Radio observations provide additional constraints as well. The early time radio observations are scintillated as the inhomogeneities in the local interstellar medium cause modulations in the radio flux density for a source whose angular size is smaller than the characteristic angular size for scintillations\footnote{1997, New Astronomy, 2, 449, 1997, Nature, 389, 261, 2008, ApJ, 683, 924}. The VLBI observations put strong size constraints as well as directly measure the speeds of the jet. Catching the early radio reverse shock (RS) emission (1-3 days) is critical in constraining the Lorentz factor. In addition, the radio observations are the most sensitive probe of the immediate environments of these mergers\footnote{E.g., 2010, ApJL, 712, L31}.

The late time optically thin spectra gives information on the particle acceleration mechanisms. Due to the slow evolution of the synchrotron in radio bands, the radio emission is detectable even when the GRB jet is in the non-relativistic regime. However, by this time, the jet geometry has become quasi-spherical. Thus, one can constrain the energetics irrespective of the information on the jet geometry. Even in cases where the jet is off-axis, only possibility of EM detection is the late time measurements at radio bands. That is because the slowing down of the jet may reveal radiation towards earth, detectable only in radio bands at late times. 
In addition, the late-time radio observations of the sub-relativistic kilonova afterglow  can complement post-merger GW searches aimed at identifying the nature of the leftover merger remnant and provide indirect constraints on the EoS. X-ray emission from this component has already been seen in case of GW 
170817\footnote{arXiv:2104.02070}, and it may
be revealing itself in radio bands quite soon\footnote{2021, ApJL 914, L20}.
Another advantage of radio observations in searching for EM counterparts is the quieter transient sky than that in optical bands, hence fewer false positives.

Radio observations are also capable of constraining some fundamental constants. For example, the joint GW plus EM analysis led to Hubble constant of $H_0=74^{+16}_{-8}  \rm km s^{-1} Mpc^{-1}$. The uncertainly in $H_0$ was dominated due to the degeneracy between the source distance and the observing angle. The radio VLABI observations put strong constraints on the observing angle which allowed to improve Hubble constant significantly, i.e., $H_0=70.3^{+5.3}_{-5.0} \rm km s^{-1} Mpc^{-1}$ \footnote{E.g., \url{https://arxiv.org/abs/1806.10596}}. It is to be noted that while 50-100 GW events needed to resolve the tension between the Planck and Cepheid–supernova measurements, only 15 GW170817-like events, having radio images and light curve data, suffice. 

With the improved sensitivity of the ngVLA working in tandem with GW observatories, one can put constraints on the radio polarization of the BNS afterglow which, together with very long baseline interferometry  (VLBI) efforts aimed at directly characterizing the structure of the ejecta, can reveal information (which so far remains inaccessible) regarding the structure of the magnetic fields. The GW-triggered events are ideal to do this because they are easier to see off-axis, hence potentially
constraining the   polarization signal  with enough sensitivity.

GW plus radio synergy will also reveal critical information on the supernova (SN) explosion mechanism. CE will be sensitive to SNe explosions within Milky Way or at most its satellite galaxies. This will be complemented by the proposed Deep Underground Neutrino Experiment (DUNE) which will also be sensitive to detect such SNe.	While radio measurements will trace the mass loss history of the progenitor and hence constrain the progenitor model, the neutrinos and GWs will unveil the heart of the star, where EM cannot reach. 

Another potential area for GW plus radio synergy is the origin of fast radio bursts (FRBs). There are currently two most favourable origins of FRBs: magnetar origin (for repeating FRBs) and the BNS merger origin (non-repeating ones). The discovery of faint pulses from the Galactic SGR 1935+2154 has confirmed that at least some FRBs may originate via the first channel.  Radio observations are most sensitive to density of the environment and can potentially identify between the two possible progenitors. However, there are two fundamentally different strategies to discover FRBs (or flares) in BNS mergers. i) Search for FRBs after GW is detected from mergers involving at least one NS. In such a case, fast response of radio telescopes is critical\footnote{2021, MNRAS, 505, 2647}. One can also rely on radio surveys. In such a case building an automated FRB detection pipeline will be most critical; and ii) to search for GW if a FRB is detected. For such cases it is important to go back to the GW database and search for an event. In both cases, accurate localization of the signal is necessary.

In conclusion, I would like to emphasize that the GW plus radio synergy has potential to bring revolution in the field of BNS and NS-BH mergers, local supernovae, probe of compact remnants of first stars to current universe and many new phenomena! The GW Astronomy is an international effort!!! People from various backgrounds, such as  physics and astronomy, engineering, computer science, data science, and statistics etc. need to come together. The most critical information needed from GW observatories is the quick access to data and the smaller localization boxes. The next generation GW observatories are capable of meeting these challenges. The upcoming radio facilities such as ngVLA and SKA are well placed to keep up with the GW efforts in this field. The radio plus GW synergy is likely to open up a new parameter space. The scope is only limited by our imagination!

\subsection{Gamma/X-ray Observations}

\textit{Presenter: Eleonora Troja}

The most obvious high-energy counterparts of GW sources are gamma-ray bursts, short-lived and bright flashes of gamma-ray emission produced by a relativistic outflow launched by the merger remnant. 
Joint GW/GRB detections probe the critical few seconds following the merger,  allowing us to explore the formation of rapidly rotating massive NSs, the birth of BHs and how they affect their surroundings by launching beams of highly energetic radiation. There are a number of questions left open by GW170817 whose gamma-ray counterpart did not resemble a standard GRB in many aspects. 
For example, the mechanism driving the gamma-ray emission (typical internal shocks or cocoon break-out?) and the origin of the delay between the GRB and GW signals (viewing effects, formation of an intermediate NS or jet launching?) remain unknown. It is likely that we will be able to address these questions not through sporadic joint detections but through a population study. 
Similarly, the elusive BBH mergers might produce weak EM counterparts, but only in rare circumstances and only a statistical study of joint detections would allow us to confirm it or rule it out. 
There are a number of approved pathfinders (e.g. HERMES), planned for launch in the next few years, to demonstrate the feasibility of a constellation of nano-satellites offering nearly complete coverage of the gamma-ray sky at a sensitivity a few times better than the Fermi GBM. Given their modest costs and technological readiness, it is conceivable to think that a similar fleet of gamma-ray satellites will be operative at the time of CE. 

An even more intriguing experiment would be to extend these prompt studies to lower energies (e.g. soft X-rays), where different components of emissions such as precursors, extended emission and magnetar plateaus are better visible. These X-ray features, although present only in a fraction of  SGRBs (~8-20\%), could be directly linked to the progenitor properties, such as its mass and spin, and provide us with novel constraints on the NS EOS before and after the merger. To enable these observations, early pre-merger warnings are a critical component.  
Several wide-field (~1000 sq deg) X-ray monitors with rapid slewing capabilities (like Swift) are under study and will be proposed to forthcoming NASA calls (e.g. SIBEX). Luckily the AstroDecadal 2020 strongly endorsed the launch of small- and medium-size missions with time-domain capabilities in the X-ray, IR and UV.  These could be launched in the late 2020s - early 2030s and have a good chance to be operative at the time of CE.  An early (~5 min) alert – even with a crude localization – would allow these X-ray imagers to repoint the GW location and stare at it right at the time of the merger, catching all the features not accessible through gamma-ray observations. 

Athena, the next flagship X-ray mission, is one assured partner for both LISA and CE-ET. Athena is planned for launch in the early 30s, with a mission lifetime of 4-5 years plus possible extensions. Given that the Decadal recommended to postpone Lynx, we consider these extensions likely as there will be no other observatory to serve the X-ray community. Athena combines rapid ToO capabilities ($\le$ 4 hours in 50\% of the cases) with high sensitivity (x10 Chandra) and will allow us to open new space for discoveries by looking at the first few hours after the merger with unprecedented sensitivity, achieving constraints that neither Swift nor Chandra can place due to their limited sensitivity and longer response time, respectively.  For additional information on Athena and its multi-messenger capabilities, read the Multi-Messenger White Paper\footnote{\url{https://ui.adsabs.harvard.edu/abs/2021arXiv211015677P}}.

\textbf{There is a wealth of science available by combining gamma and X-ray observations with GWs, but the availability, timing, and lifetime of space missions are critical ingredients to realizing the synergy.}

\subsection{Nuclear Physics}

\textit{Presenter: Andrew W. Steiner}

\textbf{It is clear that gravitational waves can provide insights into nuclear physics which are probably inaccessible otherwise.} Let's look first at some key questions which could be pursued with high-SNR GW BNS waveforms alone, complemented with advances in nuclear physics theory.

Determining the relationship between pressure and energy density at
zero temperature, $P(\varepsilon)$ over the entire range of densities
probed in neutron star interiors is possible if three conditions are
met: (i) nuclear theory can determine $P(\varepsilon)$ at low
densities (ii) gravitational wave observations can determine the
relationship between tidal deformability and gravitational mass,
$\Lambda(M)$ over all physically realizable neutron star masses, and
(iii) there is sufficient overlap between the low densities described
by theory and the high-density region constrained by $\Lambda(M)$. In
addition, future work on understanding the role which $\delta
\tilde{\Lambda}$ plays in the post-Newtonian expansion may be
required. The recent detection of gravitational wave signals from
several neutron star-black hole binaries with large mass ratios
provides some evidence that $\Lambda(M)$ could be measured across the
full mass range in the near future. \\

Going a step further, a complete specification of the equation of state (EOS) would, for example,  entail a calculation of the Helmholtz free energy
density as a function of density, temperature, and all of the
densities for the relevant degrees of freedom. While the
zero-temperature pressure-energy density relation is likely to be
well-constrained by the pre-merger gravitational wave signal,
determining the full equation of state (and all of the associated
derivatives like the specific heat, the compressibility, and the speed
of sound) from gravitational waves will require information from the
post-merger signal and merger simulations to connect the post-merger
signal to the EOS. This would argue that the 20\,km Cosmic Explorer with its targeted sensitivity in the post-merger frequency range will be important to determining the nuclear EOS.

While electromagnetic observations of neutron stars may, in general,
have larger systematic uncertainties, constraining the full EOS
will likely require combining gravitational wave information with
photon-based observations of neutron stars which are not merging. \\

At lower densities where the ground state of quantum chromodynamics consists principally of neutron and proton degrees of freedom, the equation of state is only one of a number of quantities which are
interesting for nuclear physics. In addition, it will be useful to use
mergers to constrain neutrino opacities, viscosities, the nucleon
dispersion relations and single-particle potentials, the extent of
correlations between nucleons, and the parameters of the
nucleon-nucleon interaction. Additionally, all of these quantities are
of interest in both the core and the crust, where exotic nuclei
dramatically modify transport properties. Finally, the possible
presence of non-nucleon degrees of freedom such as hyperons or quarks
is a critical question for nuclear physics and QCD -- addressing, for
example, the relationship between the dynamical breaking of chiral
symmetry and deconfinement. All of this nuclear science is potentially
accessible in the post-merger gravitational wave signal. Again, merger
simulations are required to recast the gravitational wave signal into
answers to these nuclear physics questions. Thus, all of the nuclear
physics enumerated above is also required to perform an accurate
merger simulation. \\

Determining the abundances of
r-process nuclei created in each neutron star merger is at the center
of several nuclear astrophysics questions. What fraction of r-process
elements come from mergers? Do mergers produce all r-process elements
up to the actinides? How much do r-process abundances change depending
on the nature of the merger? Answering these questions from
gravitational wave and the concomitant electromagnetic light curves
additionally requires a description of nuclear masses and nuclear
reaction rates. Some of these masses and reaction rates are measurable
in new facilities like the Facility for Rare Isotope Beams (FRIB), and some are comparatively inaccessible.
\\

{\it{Related frontiers in nuclear physics:}} \textbf{There are several important related nuclear physics questions which will need to show progress in order to fully profit from the GW and EM neutron-star observations}:

\begin{itemize}
\parskip -3pt

\item Can we improve the accuracy of results obtained from
  chiral effective theories above the saturation density,
  e.g., by calibrating the interaction to heavier nuclei or
  adding new degrees of freedom?
\item Can we accurately compute the density and energy associated with
  nuclear saturation? 
\item What is the spin response of nucleonic matter?
\item How can we improve energy density functionals to
  describe exotic nuclei and their associated nuclear
  reactions? 
\item Can new quantum many-body methods or quantum computing
  help address the fermion sign problem?
\item In what density and temperature regimes is the random phase
  approximation an accurate method for computing neutrino opacities?
  What many body methods are most effective at higher densities?
\item How do we combine our models of matter constructed with nucleon
  degrees of freedom with models of quarks and gluons at the
  quark-hadron phase transition?
\item How do we create usable nuclear physics input for merger
  simulations? We will likely, in the near future, want to go beyond
  the traditional inputs of the equation of state and neutrino
  opacities to include quantities like the nucleon single particle
  potential. However, these potentials are only uniquely defined in
  the context of a particular many-body method. Do we need to think
  about other ways of characterizing the nuclear physics input in
  terms of observable quantities?
\end{itemize}

%%%%%%%%%%%%%%%%%%%%%%%%%%%%%%

\subsection{Prospects and Challenges for Optical/Infrared Counterparts in the CE Era}

\textit{Presenter: Wen-fai Fong}

The detection of optical and infrared (optical/IR) light from a binary neutron star (BNS) or neutron star-black hole (NSBH) merger gives a wealth of information on the astrophysical outflows from mergers. The primary sources of optical/IR light are thermal transients powered by the radioactive decay of neutron-rich heavy element isotopes created in the merger (“kilonovae”) and non-thermal synchrotron emission powered by the relativistic (often jetted) outflows from the poles of the remnant (“afterglows”).

The main science drivers for studying the opt/IR counterparts are:

\begin{itemize}
\parskip -3pt

    \item Deciphering the origin of heavy (r-process) elements, and the relative contributions of mergers versus other explosive phenomena (such as Type II supernovae)
    \item Quantifying the impact of chemical enrichment from mergers on the evolution of galaxies.
    \item Building an observational picture for the interaction between the mildly relativistic kilonova ejecta and the relativistic afterglow, and the effect of that interaction on observational signatures (e.g., jet structure)
    \item In support of GW observations, placing complementary constraints on the maximum mass of NSs by inferring the nature of the central remnant left after the merger, thereby placing indirect constraints on the NS Equation of State.
\end{itemize}

\textbf{Cosmic Explorer (CE) provides an exciting opportunity to detect NSNS mergers to z$\sim$5 and beyond, allowing us to explore the optical/IR counterpart science drivers as a function of cosmic time.} This is an extremely promising avenue of research. We can also leverage the population of cosmological short-duration gamma-ray bursts (SGRBs), which are routinely localized and detected over z$\sim$0.1-2 which provide an automatic baseline for comparison to a population of CE-detected events.
\medskip

\noindent \textbf{Clear Challenges as we Move Forward in Time and Outward in Distance}
\smallskip

\textit{The Triumph of GW170817:} The discovery of light across many orders of magnitude in wavelength from the first BNS merger, GW170817, will forever be a landmark event. In particular, the rapid “blue-to-red” color evolution of its kilonova in the first ~20 days, coupled with broad features in contemporaneous optical/IR spectra, were clear-cut signatures of r-process nucleosynthesis. The synchrotron afterglow remained visible for much longer, thanks to Hubble Space Telescope Observations, through ~400-days post-merger. Coupled with X-ray and radio observations of the afterglow, the afterglow spectrum and its evolution over time was used to constrain key properties of the jet (geometry/opening angle, kinetic energy, microphysics, viewing angle).

\textit{Challenges highlighted in the status quo:} We use the example of GW190814, with component masses of $\sim$23.1\Msol and 2.59\Msol at a distance of 241 Mpc to highlight current challenges with optical/IR counterpart searches. GW190814 had the key ingredients necessary for follow-up: (1) a fairly small final localization map of 19 $\rm deg^{2}$, palatable for many optical/IR observatories, (2) prospects for an EM counterpart given the ambiguity of the mass of the second object (and potentially the first announced NSBH merger in the circulars), and (3) not extremely distant, within the ~250 Mpc horizon. Thus, the way that the community followed up this event is a fantastic proof-of-concept for what we are (in)capable of at present. Despite many observatories pointed at this event, and employing different search strategies (galaxy-targeted versus wide-field mosaic approaches), only a couple of searches actually reached the adequate depths to rule out a kilonova of similar luminosity to GW170817 and covered the entire localization region. Although it may be unsurprising to have not seen an optical counterpart (given the likely BBH merger classification), this highlights that even with our most capable resources at present-day, we are definitely resource- and instrumentation-limited at the fairly ``local” distances of LIGO/Virgo events, compared to those which will be detected by CE.

Moreover, across all O3 NS-involved events, the number of viable candidates that could still be consistent with the behavior of kilonovae (and have not yet been ruled out to date) still remains high. The advent of real-time tools, such as services which offer photometric or spectroscopic redshifts, or pre-merger detections, are fantastic resources that should be leveraged in O4, A+ and the CE era.

\medskip
\noindent \textbf{Priorities for CE}
\medskip

The light is a bit brighter at the end of the tunnel, though. \textbf{The pursuit of SGRB counterparts and afterglows have demonstrated that precise localizations are key to ever finding a counterpart}. Small-field instruments that can feasibly detect afterglows or kilonovae out to z$\sim$2 are abundant, and will become even more numerous with the advent of the ELTs during the CE era. Thus, localization is the number one priority to mitigate the need for wide-field follow-up and reduce the overall optical/IR candidate/false positive number. For reference, the average SGRB localization from Swift is a few arcseconds, and in almost all cases, a host galaxy can be identified.

\textbf{A second recommendation is to provide component masses or a firm idea of the probability of ejected material, as soon as the collaboration is able.} We will become increasingly limited by the resources capable of pursuing optical/IR counterparts at these redshifts. Thus, in the era of multiple events per week, this is crucial to inform the observational community on how we can devote our limited resources to follow the most interesting events (or more specifically, those events with high probability of having ejected material).

One other frontier to look forward to in the CE era is the advent of redshift surveys, such as DESI (spectroscopic) and Rubin Observatory’s Legacy Survey of Space and Time (LSST) (photometric) which will enable us to provide informed host galaxy associations based on \textit{a priori} information. By cross-matching the GW luminosity distance with available hosts in the localization with known redshifts, one could hope to better direct follow-up resources.

Finally, \textbf{optical observatories should be making themselves ``time-domain ready” in terms of rapid response, rapid data availability, and data reduction pipelines.} This will streamline the ability for observatories to respond in real time. The Rubin Observatory is perfectly suited to carry out this type of science, perhaps as a more ``full-time job” once the LSST survey is concluded.

%%%%%%%%%%%%%%%%%%%%%%%%%%%%%%%%

\subsection{Cosmic Neutrinos}

\textit{Presenter: Ke Fang}

Cosmic particles are a unique type of messengers that are complementary to waves in probing the universe. They include neutrinos, which are weakly interacting, neutral particles that can travel over cosmological distances, and cosmic rays, which are charged, strongly interacting particles that are confined by magnetic fields. While high-energy photons trace the footprints of both relativistic leptons and hadrons, high-energy neutrinos can only be produced by interactions of relativistic hadrons. High-energy neutrinos thus open up a clean window on nature’s hadron accelerators. 

A population of high-energy neutrinos has been detected by the IceCube Neutrino Observatory. They present a flux that cannot be explained by known models of atmospheric neutrino and muon backgrounds and therefore must be of astrophysical origin. This year, the first cosmic tau neutrino candidate and the first hint of an electron anti-neutrino were identified in the IceCube data. Several potential neutrino sources have been found, including a flaring blazar, TXS 0506+056, and an active galaxy, NGC 1068. The sources of the bulk of the astrophysical neutrinos remain unknown. 

Neutrinos and gravitational waves (GW) may be combined as a powerful tool to study compact objects. GW and thermal neutrino emission reveal the dynamics and thermodynamics conditions of a merger remnant. The extreme environment also provides promising sites for particle acceleration and production of high-energy neutrinos. For example, the merger of a neutron star (NS) binary may result in the formation of a long-lived, or indefinitely stable, millisecond magnetar remnant surrounded by a low-mass ejecta shell. Ions in the pulsar wind may be accelerated to ultra-high energies, providing a coincident source of high-energy cosmic rays and neutrinos. The high-energy neutrino signal may be detectable for individual mergers out to 100 Mpc by next-generation neutrino telescopes such as IceCube-Gen2\footnote{\url{https://www.icecube-gen2.de/index_eng.html}} and KM3Net\footnote{\url{https://www.km3net.org}}, providing clear evidence for a long-lived NS remnant, the presence of which may otherwise be challenging to identify from the gravitational waves alone. 

Another promising source of combined neutrino and gravitational waves are core-collapse supernovae. The reach of (even) CE and ET is such that events are only anticipated at a frequency of one or several per century, but the results would be rich from an observation. A GW signal from
a galactic core-collapse supernova would encode these inner dynamics, and offer vital new information. Furthermore, since thermal neutrinos emitted from the hot and dense proto-NS are also detectable, GWs provide the complementarity necessary to unlocking longstanding mysteries about the explosion mechanism. Together, GWs and neutrinos would also provide detailed information about fundamental processes that can reveal nuclear and particle physics inaccessible in the laboratory.\footnote{Astrophys. J. Lett. 876 (2019) L9 [1812.07703]; \url{https://arxiv.org/abs/2111.06990}}

Neutrino follow-up searches have been carried out by the IceCube and ANTARES collaborations for sources in the GW transient catalogs and events observed by the LIGO/Virgo O1-O3 runs. No significant association has been identified, posing limits on the isotropic high-energy neutrino energy released by the GW events. \textbf{The next-generation neutrino observatories, and CE and ET, have the potential to detect signals from a common source. Together they provide important synergistic capabilities to reveal the physics of compact cosmos.} 

%%%%%%%%%%%%%%%%%%%%%%%%%%%%%%%%%%%%%%%%%%%%%%%%%%%%

\subsection{Panel: GW-enabled Observational Science between today and 2035}

\textit{ Sathyaprakash, Discussion leader/coordinator. 
Participants: Jenny Greene, Cole Miller, Stephen Smartt, Ashley Villar}
\bigskip

We expect the current facilities to continue interleave observation with  incremental upgrades between now and  2035. The A+, and AdV+ upgrades and LIGO-India are already funded and in the  implementation phase, and others are in a conceptual phase (see \ref{sec:roadmap}). The panel focused on what can be accomplished with the existing  facilities with regard to some of the major science questions in different fields. The panel consisted of Jenny Greene who spoke on supermassive black holes, their seeds and  their growth, Cole Miller on the population of neutron stars and black holes and Stephen Smartt and Ashley Villar talking about electromagnetic follow-up of gravitational-wave transients. The panel was asked the following questions:

Questions posed to the group include

\begin{itemize}
\parskip -3pt

    \item What are the major questions and challenges in your field? 
    \item What are the requirements from GW observations to make a concrete impact  in your field between now and 2035? 
    \item What non-GW opportunities are there to make progress in your field and how  can GW observatories complement those measurements between now and  2035? 
    \item{What challenges do you think will remain to be resolved beyond 2035?}
    \item What non-GW opportunities are there to make progress in your field and how  can GW observatories complement those measurements between now and  2035? 
\end{itemize}

\textbf{Seeds of supermassive black holes and their growth:} A major question in this field is if there are black holes in the mass range [$10^2$, $10^5$] solar masses. This is the mass range in which we have no idea but they are also interesting because their presence could hint us at the physical mechanism in the formation of seed black holes that grow to become supermassive black holes. There are three pathways: (1) direct collapse models of rare, heavy seeds of $10^5$ solar masses, (2) 100 solar mass seeds, or (3) hierarchical black-hole growth at the centers of dense stellar clusters. We do not know which of these processes led to the supermassive black holes that we see in galactic cores. 

In the next decade, we will learn exactly how black holes form and evolve with the help of gravitational and electromagnetic observations. LISA should observe some $10^4$ solar-mass mergers and the mass distribution will depend on the seeding mechanism. In the 100-solar mass range LIGO and its upgrades will provide the evidence for gravitational run-away processes. LISA will also observe mergers involving extreme mass ratios providing further evidence for the growth of seeds. Heavier seeds will lead to smaller merger rates (about 10 per year) and will occur at redshifts of a few to z~15. Lighter seeds will lead to merger rates that are an order of magnitude larger. 

The Extremely Large Telescopes including the the European Southern Observatory's ELT, and the two proposed US-led Thirty Meter and Giant Magellan Telescopes will allow us to get number densities and the dynamics, while gravitational-wave observations will help us get a combination of the number of densities and merger rates of black holes up to redshifts of z=10. We don’t know how often intermediate mass black holes merge due to the final parsec problems. Next generation x-ray telescopes will reveal the feeding mechanisms and the growth of massive black holes. Tidal disruption events will provide additional information about the growth of black holes. 

\textbf{Neutron Star and Black Hole Populations:} The major questions and challenges are understanding of the equilibrium properties of the cold dense matter in neutron star cores and the origin and diversity of the masses, mass  ratios, spins (including inclinations), and eccentricities of binary neutron stars and black holes. In this regard, GW observations will be helpful in the accumulation of data; trends and exceptions in the observed population. 
For nuclear physics, greater precision at high frequency will be extremely useful in observing the post-merger signal from binary neutron stars and binary black holes as that will be critical to inferring the dense matter equation of state and testing general relativity, respectively. For binary populations, greater sensitivity at low frequencies will help in greater accuracy in the measurement of parameters as also discovering new populations of intermediate mass black holes. 
Further progress in the field would need theoretical progress in waveform models, especially at high frequencies to understand what signatures of the dense matter equation-of-state and quark-deconfining phase transitions might be present in the post-merger signal. Furthermore, continued development of models of EM afterglow in the case of double neutron star and neutron star-black hole binaries would be necessary to interpret the data to learn about the central engines of gamma-ray bursts and the associated mechanism behind relativistic jets as also to interpret the kilonova light curves. Transition to genuinely predictive models of binaries, which means a clear  indication of what outcomes are impossible in each model would be desirable. 
Challenges that will remain past 2035 include low tidal deformability of neutron stars of masses greater than about 2 solar masses, so the NICER  radius measurement of a 2.1\Msol star may still be state of the art. 2. For populations, M$>$upper limit of upper mass gap would be wonderful but  Nature may not be that kind. Improvement in low-frequency sensitivity would  prepare us to receive Nature’s kindness. 

\textbf{EM Follow-up of GW Events:}  Two major challenges in the electromagnetic follow-up of gravitational-wave events are in the development of optimal search algorithms for the detection of post-merger prompt emissions with minimal survey interruption. First, survey times available in large telescopes will be limited and hence it is important to develop algorithms that make the best use of the allocated time. In this regard, it is important to compile complete galaxy catalogs that are relevant to the sensitivities of GW detector upgrades and new observatories. Secondly, truly multimessenger astronomy analysis methods are currently lacking as they don’t address more than two data sets at any one time. Full forward modeling of all EM  signatures would be necessary to realize the full potential of multimessenger astronomy. \textbf{EM follow-up would greatly benefit from smaller, accurate and rapidly available localization maps. However, it is particularly important to ensure that the smaller maps are not systematically offset and are accurate when first released keeping in mind that observing times on large telescopes are too expensive.} It would also be helpful to have rapid classification (e.g., `EM bright’) of detected events as well as quick updates to the classifications of the GW sources when they change.

Currently, there are early models for kilonova and afterglows but with improved observations and greater number and variety of events it is necessary to have better EM models for kilonova light curves and afterglows as well as tools ready to describe exotic phenomena. For example, kilonova would need nebular spectral models, especially to be used as calibration for early models as well as extended grids of numerical simulations. Additionally, we would need better treatment of off-axis jets to observe afterglows as there will be significant increase in the number of off-axis systems that will be observed in the coming years. Continued monitoring of  SGRBs and their hosts will be necessary to build a comprehensive understanding of the physics of these relativistic sources. 
By 2035, gravitational-wave detectors will begin to observe binary neutron star events from redshifts of 0.5 and beyond. We need EM follow-up strategies for such high-z EM counterparts and none exists today. Ensuring that Rubin Observatory is operational and running, after the 10 yr survey is complete would be critical for following up GW events both in high numbers but also at great distances. It is also crucial to have X-ray and gamma-ray detectors available on the timescale of the next two decades. Specifically, planning would need to start now to have space detectors like Swift in place by 2035.

\textbf{Open Discussion:} \textbf{Many EM observers are more interested in the continuous operation of the near-term GW network than upgrades in sensitivity, once GW detectors are sensitive to sources at z$\sim$0.2.} There aren’t any telescopes that can observe EM counterparts at greater distances and so upgrades will not help in following-up more events. At O3 sensitivities, sky localization is pretty poor except for the closest events and the redshift reach is about 120 Mpc for binary neutron stars. This will improve to 160-190 Mpc for O4; still, only a handful of events will be localized to within 20 $\rm deg^2$. In O5 the distance reach for BNS will be about 300 Mpc, z$\sim$0.07 and $\sim$10 events per year with localization typically of 20 $\rm deg^2$. In order to reach better localization than 20 $\rm deg^2$ for about 50 events per year would require sensitivities that are better than what’s expected for O5. The continuous incremental improvement will be welcome for event rates (with the gain as the cube of the sensitivity), and the approximate factor of 2 gain in SNR for a given source will also help in extracting additional information from the waveforms.
It is important to have the detectors running continuously in order to catch once-in-a-lifetime opportunities such as a supernova in our own galaxy. The answer is to have an operating network so that it will be possible to have at least some of the detectors operating at any one time. We are not there yet but could get there by the turn of the decade. 
GW detectors can be tuned to work better at higher frequencies at the cost of worsened sensitivity at lower frequencies. Should one go after guaranteed science or take a leap of faith and look for new sources? At low frequencies one is making a bet that there are sources at those frequencies but this is not guaranteed since the mass range will be very large. 100 solar-mass events might be the top end of the mass range from stellar processes but there might be nothing at 1000 solar masses. At high frequencies, $\sim$1\,kHz matters more than 3\,kHz. At 1\,kHz neutron stars will be simple: relatively cold compared to their Fermi energy, not rotating much let alone differentially rotating and information you get can be interpreted in well-understood theoretical framework modulo improvement in waveforms. At higher frequencies neutron stars will be far more complex: hot neutron stars with temperatures comparable to Fermi energy, differential rotating, neutrino viscosity. It will be a greater challenge to interpret the data and waveforms are difficult to model.  However, stellar mass black holes at high redshifts will appear as intermediate-mass black holes and hence it is still important to have good low-frequency sensitivity. 
For early warning triggers, how early should the alerts be?  For successful EM follow-up what is important is to provide accurate information about masses in addition to sky position; early warning is not important for IR as there is no rapid evolution in the IR spectrum in the first few minutes. However, optical flash and gamma rays are important and they would change on the time-scale of seconds and radio emissions could occur even before the merger.

\subsection{Next-generation Observatory science goals}

\textit{Presenter: Joshua Smith}
\bigskip

The GWIC `3G' Science 
Book\footnote{\url{https://gwic.ligo.org/3Gsubcomm/docs/GWIC_3G_Science_Book.pdf}}, 
prepared by a community-wide team of interested scientists, has formed the basis for the current models of science objectives for CE and ET. Each of those projects then considered the trades between the observational science goals and the technologies that would be needed to realize them. 

In the case of CE, a new trade study was undertaken to help focus the design. A top-level set of science goals was identified, informed by the GWIC Science Book and many discussions with the scientific community. Networks were studied consisting of one or two CE detectors, with or without ET and also considering a detector in the southern hemisphere. Parameters of the designs were adjusted to find the best combination of features to meet the observational science goals for different networks, leading to the design laid out in the Horizon Study. 

The CE team chose to identify a limited number of key objectives\footnote{Please see the Horizon Study for more detail, \url{https://dcc.cosmicexplorer.org/CE-P2100003/public}}:
\begin{itemize}
    \item \textit{Black Holes and Neutron Stars Throughout Cosmic Time:} Cosmic Explorer is designed to detect gravitational waves from black holes and neutron stars in binaries to redshifts of $\sim$10 and above. The goal is to shed light on Population III stars through the black holes they might have left behind, to measure the properties of the first black holes and their role in forming supermassive black holes and galaxies, and to characterize the populations of compact objects and their evolution.
    \item \textit{Dynamics of Dense Matter:} The Cosmic Explorer design has the capability to measure gravitational radiation from binary neutron star coalescences and provide the precise source localizations required for multimessenger astronomy. Here the goal is to determine the internal structure and composition of neutron stars, explore new regions in the phase diagram of quantum chromodynamics, map heavy element nucleosynthesis in the universe through counterpart kilonovae and distant mergers, and reveal the central engine for the highly relativistic jets that power short gamma-ray bursts. Observations of binary coalescence, and searches for continuous-wave signals from pulsars, will inform this science. 
    \item \textit{Extreme Gravity and Fundamental Physics:} Cosmic Explorer’s significant increase in sensitivity is intended to deliver exceptionally high resolution of strong gravitational waves and to increase the reach to see weak and distant sources. This can reveal the (potentially new) physics of the most extreme gravity in the universe, enabling investigation of the nature of strong gravity with unprecedented fidelity, discovery of unanticipated compact objects beyond the reach of current detectors, and a path independent of electromagnetic and particle data to pursue the nature of dark matter and dark energy.
    \item \textit{Discovery potential:} This is both the least-well defined goal of observations in this new domain, and perhaps also the most exciting. For the previous targets, modeling can lead to specific measures of the ability of a design to deliver the desired science. For discovery science, pushing the key top-level performance measures of best sensitivity, bandwidth, and numbers and placement of the nodes of the network is indicated. Cosmic Explorer and Einstein Telescope have pursued this goal in their designs, and \textit{possible} discoveries in quantum gravity, new particles and fields, and any one or more of sources of stochastic backgrounds are some outcomes currently imagined. Nature may have others.
\end{itemize}

\textbf{The observational science objectives of these next-generation observatories will continue to evolve in the coming years, informed by input from the non-GW community, theory, and the growing body of gravitational-wave signals from the current detectors.} The Observational Science presentations and discussions on the first day of Dawn VI -- and the discussions to follow -- will be significant input for the next round of design refinement for CE and ET, and the growing level of activity in planning the multi-messenger activities in instrumentation and observational science.

\subsection{Observational Science Recommendations}

Recommendations and key observations from the presentations and discussions:

\begin{itemize}

	\item A network of detectors is important to realizing the science goals of the community. This point came up throughout the meeting, and the strong consensus is that three (or more) detectors must be the goal. 
	\item Explicit coordination of GW observation with (in particular) space missions (worldwide) due to their limited lifetime is needed to ensure that MMA can be best realized.
	\item In general, there would be value in establishing a structured regular interaction between the GW and EM/particle/nuclear domains to ensure insights which can impact any of the domains are communicated clearly and in a timely way.  A small group that has visibility and credibility with all communities would be best.
	\item The desired precision of localization by a GW network, and the need for low latency vs. high precision, may merit a focused meeting or exchange. This could help drive detector designs and observatory placement (which have of course also other constraints: Technical capability and cost).
	\item `Followup Fatigue' is starting to appear. GW alerts from this time forward may have to carry additional information to aid in triage. 
	\item The tension between observing and improving the sensitivity with its concomitant downtime continues. This would be a valuable topic for a Community Town Meeting, both to look carefully at the upgrade argument and to hear more about the value of extended observation at a fixed sensitivity. 
	\item Related is the need for proper coordination between the detectors. Once KAGRA and LIGO-India at a sensitivity level comparable to LIGO and Virgo, phased upgrades will be possible. These should have input from the non-GW observing community and be aware of limited-lifetime space missions.
	\item Every effort should be made to avoid a long `upgrade' gap. Both LIGO and Virgo intend, if funding allows it, to continue to observe until CE and ET are ready to start observations.

\end{itemize}

%%%%%%%%%%%%%%%%%%%%%%%%%%%%%%%%%%%%%%%%
% detector to next generation detectors 
\section{Current and future Observatories}\label{sec:design}
\subsection{Introduction}

There has been very significant progress fulfilling many of the recommendations for the ground-based detectors developed in Dawn IV and Dawn V. The LIGO A+ and Virgo AdV+ upgrades have been designed, funded, and are largely implemented. Planning exercises in LIGO and Virgo are underway to consider the interval from between the A+/AdV+ epoch to the CE/ET epoch. LIGO-India has a site and significant R\&D infrastructure has been created. KAGRA has undertaken an initial observational run. The `3G Reports' were completed, and a more general GWIC Roadmap published. White Papers for the US Astro2020 Decadal were written and submitted. From the perspective of the next-generation observatories, this culminated in the successful proposal to ESFRI from ET, and the completion of the CE Horizon Study. This session of Dawn looked at the state of the current and planned Observatories, with a focus on the US Program. 

\subsection{The roadmap between current observatories and next-generation detectors}\label{sec:roadmap}

\textit{Peter Fritschel, discussion leader/coordinator; 
Participants: Daniel Sigg, Brian Lantz, David Ottaway, Paul Lasky, Matteo Barsuglia, Rana  Adhikari, Ryan Wollaeger}

\bigskip

The `A+' project\footnote{\url{https://arxiv.org/abs/1410.5882}, \url{https://arxiv.org/abs/1304.0670}} and the Virgo `AdV+' project\footnote{Proc. SPIE 11445, Ground-based and Airborne Telescopes VIII, 1144511 (13 December 2020); \url{https://doi.org/10.1117/12.2565418}} upgrades for 2G detectors, now in progress, 
follow an attractive incremental model: rapidly and 
opportunistically applying select new technology, as it becomes 
available, to escalate the observing horizon in stages. The O4 observing run is expected to start in late 2022 of about one year duration, with a significant fraction of the `plus' technologies installed and commissioned. The O5 observing run, with the LIGO detectors ultimately operating at the design sensitivity
targeted for A+, is currently expected to run through mid-2028.

\textbf{Beyond 2028, the LIGO Laboratory is firmly committed to continued observations of the gravitational-wave sky.} Plans are currently being developed by LIGO and the LSC (and also in Virgo) to chart a path for interferometer upgrades and observing periods beyond O5 and  improve the detectors’ strain sensitivity across the frequency band (from $\sim$10 Hz to $\sim$3 kHz employing some of the technical advances being planned for Cosmic Explorer. These enhancements can lay firm groundwork for future 3G designs, vetting  aspects of the new technology in the most stringent, realistic context currently possible. Key questions being pursued are
\begin{itemize}
\parskip -3pt
    \item What dates bracket when O5 observing will be brought to a close, and the start of a post-O5 observing run? As one observational-science criterion, O5 should continue long enough to exceed the event count had the instrument simply continued with O4 sensitivity. There are also observational-science motivation for a continuous observation. An upgrade should be `ready' before ending observation, in order to minimize downtime; \textbf{the durations of downtime and the post-O5 run should be such that the observational science goals of the greater community are best satisfied.} 
    \item What observational science drivers are there for both the kind of upgrade and the timing of the upgrade? As we learn more from observing we can move more from `the best instrument we can build' to `the instrument that best delivers the science objectives'. We also will want to try to synchronize with constrained observing programs in the electromagnetic and particle fields. 
    \item What technologies suitable for upgrades have the greatest impact on informing CE designs? Cryogenics and crystalline coatings are two examples of technologies with great potential for next-generation observatories; both come with capital costs and additional schedule, cost, and technical risks. 
    \item \textbf{The current detectors have significant excess technical noise at low frequencies, impacting current and future observational science goals. How can the designs for post-O5 detectors best address this excess (or at least facilitate the effort to identify and mitigate the excess)?}
\end{itemize}

At low frequencies, the kinds of improvements under consideration for the 3- and 4km infrastructures include larger test masses and improved suspensions for them, and means 
to reduce the motion of the seismic isolation platforms. At low-to-mid frequencies, we can employ incremental or qualitatively different coatings with lower thermal noise. At all frequencies, there is room to reduce quantum noise through higher levels of vacuum squeezing and higher laser power, again as envisaged for Cosmic Explorer.
These plans will be developed and refined in 2022 and 2023 and may further evolve in response to inputs from the experience of commissioning and operating the upgraded instruments, from the LIGO, Virgo, and KAGRA Collaborations, and from the greater observing community.

The teams looking at the upgrade path for LIGO are considering a range of options, in terms of magnitude of change and magnitude of sensitivity improvements. Changes in the system which leave the wavelength at the present 1064\,nm (Nd:YAG lasers), the main optics of fused silica, and the system at room temperature form one `branch' of possible improvements.\footnote{Employing some of the concepts explored in \url{https://arxiv.org/abs/1712.05417}} An alternative `branch' is to consider mild cryogenics, and a shift to longer wavelengths to accommodate silicon main optics.\footnote{The Voyager concept is a well-developed package of upgrades; \url{https://arxiv.org/abs/2001.11173}} In addition to technical feasibility, the downtime, cost, risk, and astrophysical reach are all important considerations. The utility as pathfinders for CE is, as well. 

The tension between observing and upgrading the instruments was discussed at several points in the meeting. In terms of numbers of events, a pause of a year or two can be more than recovered due to the growth in event rate as the cube of the detector sensitivity. There is also a gain in SNR for a given event, and some improvement in localization. However, the gaps in GW observations run a risk of missing a rare event and can deprive a short-lived EM/particle mission of significant common time with GW observation. The bigger picture includes the KAGRA and LIGO-India detectors. The dates by which those detectors can achieve LIGO- or Virgo-like sensitivities are still in flux. However, once there is a network of 5 detectors, one (or two) detectors can be taken offline for improvements while retaining a network with the capability of localization. 

The greatest challenge in reaching the target sensitivity is in the frequencies below roughly 50 Hz. Here, many different noise sources, some related and some independent, contribute to an excess above the `fundamental' noise sources of thermal noise (in the coatings and ultimately suspensions) and quantum radiation pressure noise. This domain has always proven to be difficult, and advances in the sensitivity often require in-vacuum modifications or significant changes to sensors and servo-control systems. This domain is also central to the observational science goals, and not only for the current detectors -- also for CE and ET. This is one of the motivations in the instrument science community for continuing to pursue upgrades; they provide an opportunity to learn about the limitations of the instruments, and thus to improve the designs for the following phases.

An important additional concern is maintaining a sizable team of instrument scientists who have experience with the full range of activities: identifying weaknesses in the current designs, undertaking R\&D to develop approaches to move forward, being part of the engineering, fabrication, assembly, test, and installation; and then the commissioning of the completed instrument. This gives a strong incentive to maintain a rhythm of upgrades that is roughly one `PhD cycle' in length.

\textbf{The best path for upgrades of the detectors in the current 3- and 4\,km observatories will evolve most significantly with the time scales for realizing CE and ET. }If those next-generation-observatories follow the fastest path to realization that seems feasible, more modest upgrades of the current detectors will be best. If there are delays in CE and ET, more ambitious programs will be motivated.

Pedro Marronetti of the NSF reviewed the range of options for NSF support of upgrades. There are proposal opportunities ranging from one to hundreds of millions of dollars. Evidently one of the considerations in the US is the overall availability of large-scale funding; a very significant upgrade to the current LIGO observatories may well delay the start of funding of CE.

\subsection{Einstein Telescope Project}

Einstein Telescope\footnote{\url{http://www.et-gw.eu}} is a pan-European concept for a next-generation gravitational-wave observatory. 

\subsubsection{ET Concept}

The ET concept\footnote{\url{https://gwic.ligo.org/3Gsubcomm/docs/ET-0007B-20_ETDesignReportUpdate2020.pdf}} is based on a triangular array with 10\,km long arms and three co-located detectors. Each detector consists of two interferometers in a xylophone configuration, where one interferometer maximises sensitivity at low frequencies while the other interferometer maximises high frequency performance. The output signals from the two interferometers are then combined to provide broadband sensitivity for each detector. The triangular system has the ability to resolve the polarisation of gravitational waves without additional GW observatories, and the 6-interferometer implementation makes a very good use of the underground facility. In addition, a combination of the output signals of all three co-located detectors (null stream) allows a validity check of detections due to the redundancy of the signals and thus a slight reduction of the detection threshold, which is equivalent to an increase in sensitivity.  The main challenge is the complexity of installing, commissioning and operating multiple instruments in the same facility.

The xylophone configuration allows the decoupling of technologies that are difficult to coexist in a single detector, such as cryogenics ($\sim$20K) and high power operation. The ET-Xylophone concept envisages a cryogenic low-frequency detector with low circulating power (18\,kW) and a high-frequency detector at room temperature with high circulating power (3\,MW).  This split into two interferometers, with partly very different requirements, increases the complexity and the commissioning effort. Experience gained on one interferometer cannot necessarily be transferred to the other type.

ET is built underground at a depth of about 200-300\,m to minimize seismic disturbances and the gravitational interaction of the mirrors with their environment (Newtonian Noise).
This requires an underground infrastructure with a total of about 30\,km of tunnels and various underground chambers with heights up to 30\,m.

\subsubsection{Project Status}

As a pan-European project, one of the main organizational challenges is the coordination of activities at the political level. ET was placed on various national roadmaps. Most significantly, with the political support of Italy, the Netherlands, Belgium, Spain and Poland, \textbf{ET was included on the ESFRI roadmap in June 2021, which is equivalent to a quality label at the European level and an important milestone for ET.}
The internal organization of ET is being advanced at both the project and the collaboration level, on the one hand by establishing the ET project with the support of the involved funding agencies and by the formalization of the collaboration on the other hand. The collaboration already includes several well-functioning specific boards that take care of the instrumental aspects (Instrument Science Board, ISB), the scientific output of the project (observational Science Board, OSB), the survey of the site candidates (Site preparation board, SPB), the computing needed for the data analysis (Computational Infrastructure board, CIB), and various others.
There are currently two candidate sites for the observatory: the Italian island of Sardinia and an area called Euregio Meuse-Rhine, the border region between the Netherlands, Belgium and Germany. Studies are underway at both sites to explore their scientific suitability in terms of seismic disturbances and subsurface geology. The site decision is envisaged in 2024 or 2025.
The next few years will be characterized by the so-called Preparatory Phase of the ET project, during which the organizational structure of ET will be established, the technical studies will be refined and deepened, the site decision will be prepared both on the scientific and on the policy level, and funding will be pursued.
The projected timeline foresees a start of construction in 2026 with a construction phase lasting up to 8 years. After the installation of the vacuum system and the first detector, a first data taking can be expected in the second half of the thirties.

At present, formal Coordinators of the ET Project are INFN in Italy and Nikhef in the Netherlands. A Board of Governmental Representatives (BGR) has been established with ministerial delegates from various interested countries. The Coordinators report to the BGR. In addition a Board of Scientific Representatives (BSR) has been established that includes scientists from various countries. Moreover the Coordinators have set up an ET Project Directorate. The ET Project is now in the Preparatory phase and important deliverables of this phase include the agreement on the type of legal entity, governance, project organization, technical design, financial planning, and site selection. 
\subsection{NEMO}

\textit{Presenter: Bram Slagmolen}

The Neutron star Extreme Matter Observatory (NEMO\footnote{Publications of the Astronomical Society of Australia, 37, E047. doi:10.1017/pasa.2020.39}) is a concept for an interferometric gravitational wave detector focused on observing the late in-spiral and post-merger signatures of Binary Neutron Stars to infer their equation of state, in an infrastructure of $\sim$4\,km. To have a realistic chance of observing these signatures a sensitivity window reaching at least $10^{-24} \rm Hz^{-1/2}$ in a kHz bandwidth centred near 2 kHz is needed. This would be a comparable sensitivity in this frequency range to other next-generation detectors such as Cosmic Explorer and the Einstein Telescope but would be achieved with a specialized detector.

The sensitivity requirements at the low-frequencies ($<$ 1kHz) are relaxed to enable high bandwidth control loops to mitigate opto-mechanical instabilities and reduce the cost with less complex test mass suspension systems. The planned optical configuration is similar to the current observatories using a dual-recycled Fabry-Perot Michelson interferometer. Compared to other detectors however, it will employ a long signal recycling cavity to tune and maximise its sensitivity in the 1–4\,kHz band. In addition, alternative signal enhancement techniques can also be utilised to further improve the target sensitivity.\footnote{Classical and Quantum Gravity, 37(7):07LT02, Mar 2020; Communications Physics, 4(1):27, 2021}

NEMO is envisioned to employ evolution of present technologies, such as the use of longer laser wavelengths\footnote{Opt. Express, 28(3):3280–3288, Feb 2020} and cryogenically cooled silicon test masses\footnote{Phys. Rev. D 102, 122003. doi: 10.1103/PhysRevD.102.122003}. This will make a NEMO detector a pathfinder for technology that will enhance the next generation detectors. A NEMO pathfinder is supported by the Australian Astronomy Decadal Plan\footnote{\url{https://arxiv.org/abs/1912.06305}}, and a conceptual design proposal is currently in preparation.

\textbf{More generally, the observational science value to having a network node in the southern hemisphere is significant. The community should continue to explore means to realize a next-generation observatory there.} 
\subsection{The Cosmic Explorer Project}

A series of three presentations provided a view of the current state of Cosmic Explorer and the path forward.

\subsubsection{CE Concept and Status}

\textit{Presenter: Joshua Smith} 

The US Cosmic Explorer concept is an outgrowth of studies performed by members of the GW instrument science community in parametric explorations of concepts for future detectors. The community participated over many years, via the GWADW, GWPAW, and PAX workshops, the Amaldi meeetings, and the LSC working groups and meetings, to explore the paths for significant improvements to the detectors. Interaction in a variety of environments and covering both observational science goals and instrument science paths to realization has led to the specific concept of CE. The name ``Cosmic Explorer'' with the nominal length ten times that of LIGO was described in the 2014 LSC Instrument White Paper,\footnote{\url{https://dcc.ligo.org/LIGO-T1400316}} and a growing level and focus of activity has continued since that time to find the best path forward. 

\textbf{Guided by the experience with the LIGO and Virgo detector commissioning, the CE team came to the conclusion that while making the detector longer evidently increased the cost, it appeared to be the lowest risk path to better sensitivity.} Initial comparisons with observational science targets suggested that the compromise of building a surface detector -- which would be limited at low frequencies by Newtonian noise coupling to seismic activity -- would preserve, and deliver, a wide range of exciting science targets; the great length also suppresses the size of this noise source. There is an explicit understanding that multiple generations of detectors, which can push the sensitivity further, will be accommodated over the lifetime of the observatories. With that starting point,\footnote{\url{https://arxiv.org/abs/1607.08697}} the Cosmic Explorer Project could be proposed to the NSF for a serious pre-conceptual-design phase study,\footnote{NSF Award \#1836814, Collaborative Research: The Next Generation of Gravitational Wave Detectors} leading to the recently completed Cosmic Explorer Horizon Study. 

The CEHS was carried out by a group of five institutions.\footnote{MIT, Caltech, Cal State Fullerton, Penn State, Syracuse} Faculty and research scientists, postdocs, grad and undergrad students contributed directly to the focused research and writing. Civil engineering support was contracted. The LIGO Lab and OzGrav, along with members of the gravitational-wave community broadly, helped to form, critique, and share the concept. A CE Consortium has been formed\footnote{\url{http://cosmicexplorer.org}} to focus this support to future work. 

 \textbf{Cosmic Explorer is envisioned to be the next-generation US-led observatory project consisting of two widely separated and non-parallel sites; one site supports a detector with 20\,km long arms,  and the other a detector with 40\,km long arms.} The 40\,km system is designed to deliver 10 times better sensitivity than the 4\,\,km LIGO A+ design, with a primary observational science goal of access to most neutron star and (stellar to intermediate mass) black hole mergers in the universe. The 20\,\,km detector is optimized for neutron star post merger physics; its shorter length allows an optimized sensitivity in the range of 1-3\,kHz. The two would observe together, ideally with ET and conceivably with a southern hemisphere detector, forming the next-generation gravitational-wave detector network.
 
 The two sites would use common technologies and fabrication processes for all of the civil and vacuum infrastructure, and the vast majority of the detector optical, mechanical, and electronic components. The experience with LIGO's two sites demonstrated the utility of commonality for  cost savings in construction, and the advantages for commissioning of solving problems once and training staff who can work interchangeably at the two sites.
 
An initial `chalkboard' budget for the CE project has been developed. The costs of the design and construction phases are based on scaling from Initial LIGO, Advanced LIGO, and engineering estimates made specifically for CE. \textbf{The cost for construction of the two sites and the detectors for them is roughly estimated at a cost of \$1.6B 2021 USD. Operations then follows, with a yearly cost estimated to be \$60M 2021 USD.} It must be stressed that they are a first effort and require refinement in the next phase of the Project. 

\subsubsection{Cosmic Explorer Reference Design}

\textit{Presenter: Stefan Ballmer}

\bigskip

The Cosmic Explorer reference design calls for a 40\,km and a 20\,km L-shaped surface-based gravitational-wave detector. The design is guided by the science goals outlined in the CE Horizon Study, capable of observing frequencies up to about 5\,kHz from neutron star mergers, and mergers of black holes out to high redshifts ($\sim$10).
Notably, the characteristic strain from binary black hole mergers at a reference frequency of 10Hz  exhibits a plateau behavior for redshifts above a few, due to a combination the characteristics of the luminosity-distance -- redshift relation and the frequency shifting of stronger-emission late-inspiral signals. The signal will drop away once the merger frequency has been shifted below the reference frequency. The result is a “threshold sensitivity” for reaching cosmological distances with gravitational-wave interferometers.

\textbf{The CE reference design is guided by three principles: i) “Build on what works”, taking advantage of the success of Advanced LIGO’s design as much as possible; ii) “Let observational science drive the design”, i.e., match the antenna length to known sources and address the known issues with Advanced LIGO’s design; and iii) “Keep it flexible”, designing an observatory that will be able to accommodate any new technology developments.}

Other than the increase in arm length (from 4\,km to 40\,km/20\,km, including larger vacuum system), the reference design asks for larger test masses (320\,kg, 70\,cm diameter, compared to Advanced LIGO’s 40\,kg, 34\,cm diameter), a second input mode cleaner to address the frequency stabilization requirements, beam reduction telescopes on the arm-side of the beam splitter, a low-loss signal recycling cavity, a scaled up filter cavity to rotate the squeezed vacuum for quantum noise suppression, and a homodyne readout scheme.

The required R\&D is outlined in much more detail in the CE R\&D white paper,\footnote{\url{https://dcc.cosmicexplorer.org/CE-P2100005}} as well as the CEHS.\footnote{\url{https://dcc.cosmicexplorer.org/public/0163/P2100003/007/ce-horizon-study.pdf}}

\subsubsection{ CE Path Forward}

\textit{Presenter: Matt Evans} 

\paragraph{Cosmic Explorer Timeline mapped onto the NSF Major Facilities Lifecycle}

The CE team is using the NSF Major Facilities Guide to structure the work going forward. Evidently if other funding agencies can participate in the project, additional direction will come from the practice in those agencies. The Cosmic Explorer Horizon Study (CEHS) fits well as a conclusion to the NSF `Development Phase', ``in which initial ideas emerge and a broad consensus is built for the potential long-term needs, priorities, and general requirements”. With the support of GW future science from the Astro2020 Decadal and for Cosmic Explorer from the Dawn meeting, we anticipate NSF recommendation to enter the `Design Stage' “where detailed, construction-ready budget estimates, schedules, technical specifications and drawings, and management processes are developed”. 

The presentation followed by a day the ``Megaprojects'' round table as well as the other Dawn meeting sessions, and an effort was made to incorporate some of the ideas from those presentations and discussion in what follows (and what will be planned for the future).

A sketch of the plan forward is laid out in the CEHS which calls for an overall Design phase of around 7-9 years at cost of order \$100M 2021 USD:
\begin{itemize}
\parskip -3pt
    \item Conceptual Design 3 years
    \item Preliminary Design 2 years
    \item Final Design 2-4 years
\end{itemize}
This will be followed by a Construction phase of some 5 years, and the Operations. The infrastructure will be designed for a lifetime of order 50 years. This will allow funding agencies to capitalize on future research and development breakthroughs, should the operational life-span
of CE be extended beyond the initial mandate (which is expected to be 20 years).

The baseline plan for obtaining funding for the near-term activities of Conceptual Design is for a multi-institution proposal to the NSF be made which would support a Project office, enable an initial Project Execution Plan to be written, and to work intensively to identify a `home' for Cosmic Explorer in an organization capable of managing this scale of project. In parallel, R\&D activities in the broader community will be coordinated by the Project Office but supported by NSF (and ideally other US and international funding agencies), in a mode resembling Advanced LIGO in its coordination of the nascent LIGO Scientific Collaboration R\&D. 

\paragraph{Core Activities along the path to realization}

We outline the work to be done according to the NSF Major Facility Guide (MFG) phases.

\subparagraph{Conceptual Design Phase}

The development of the Project governance, schedule, and budget will be the lead activity in this phase. Identification of a structure, and likely a host organization, for the Project is critical. 

The observational science goals will continue to be refined, informed by the greater scientific community's research into multi-messenger applications of GWs, and the growing body of GW observations. The goals will be flowed down to the requirements for Cosmic Explorer.

The identification of site candidates will be another focus in this phase. Impact assessments and acquisition feasibility will be pursued. Civil engineering will investigate the Newtonian noise and site preparation plans both generically and for target sites. Parametric estimates of the cost will be developed with substantial engineering backup. 

\textbf{Inextricably linked with the site search for CE is the need to build relationships with the Indigenous Peoples to whom the land belongs.} These relationships will inform the CE team's approach to identifying suitable sites, and how one could proceed with integrating CE into the land in a way which is not just compatible with the terrain and culture but forms a partnership that is productive and positive for all the parties involved. Effort will be made to avoid barriers -- physical, organizational, or philosophical -- between CE and the environment. 

\textbf{The vacuum system for CE is the largest cost element and also one that impacts the implementation approach.} Prototype tests and modeling will be undertaken, and again parametric cost estimates developed. 

The initial detector designs will be refined. Insights from the continued work with the current detectors will be incorporated, and where feasible implementation of Cosmic Explorer designs will be integrated into LIGO for testing. The flowdown of observational science goals will influence the priorities in the design. Seismic isolation and thermal noise considerations will drive the isolation and suspension systems. Mirror coatings are likely to drive R\&D, with crystalline coatings one path to explore. Interferometer layout and control will be detailed.

\subparagraph{Preliminary Design Phase}

The Preliminary Design will iterate on the Conceptual Design phase products. The site selection, anticipated to be an NSF activity, will be supported by the CE team; Indigenous Peoples will continue to play a key role. The design options for the detector, vacuum system, and civil construction will be down-selected to the optimal subset. A bottom-up cost estimate will be made, risks assessed, and a detailed Project Execution Plan (PEP) developed. 

A transition to the target Project management organization will be completed during this phase.

\subparagraph{Final Design Phase}

The relationships of the now-firm Cosmic Explorer Observatory management with the scientific community will be established, and plans crafted to create a sustainable and diverse team and community. The CE presence at the sites will grow as the team is integrated into the local communities, and the local communities engaged in the CE Project. The PEP will be finalized. Vendors for the key items will be identified and the logistics of the construction process determined. 

At the close of this process, the Project should be well underway and the installation, commissioning, and operations activities will become the new focus.
\subsection{Round Table on Data Access Models}

\textit{Duncan Brown (chair), Edo Berger, Jenne Driggers, Andreas Freise, Jeff Kissel, Alex Nitz}

\bigskip

\textbf{Data Analysis:} The computational cost for compact-binary searches is manageable in Cosmic Explorer as the cost scales principally with the bandwidth of the detector, not the overall sensitivity. Cosmic Explorer’s increased bandwidth requires only 2-3x more computational power than Advanced LIGO. However, there are a number of technical challenges facing compact-binary searches that will need research and development over the coming years. These include: power-spectral density and background estimation in the presence of multiple signals, improved methods for accessing low-frequency sensitivity. Searches for continuous wave signals remain the most expensive and will require continued investment into methods to make the most of available computing resources. The substantial data-analysis challenges are associated with parameter measurement, given the expected signal rate. The Cosmic Explorer challenge differs from LISA; the Cosmic Explorer data set is much larger and signals are expected at a rate of 1/minute, and so some overlap can be anticipated, but are generally not confusion limited. Fast methods of parameter estimation that amortize the costs (e.g., machine learning) or provide rapid computation of the likelihood (e.g. reduced-order or heterodyne likelihoods) will need technical development. A significant effort will be required to automate signal detection, parameter measurement, and validation to cope with the high rate of signals. Extracting the science from loud signals will require much more accurate waveform models than currently available. Continued investment in numerical simulations and supporting theory (especially nuclear theory for neutron star mergers) will be essential.

\textbf{Data Production, Calibration, and Delivery:} A key element in achieving success in CE era must be a paradigm shift in human resources for these corners of operation. To date, these activities have only had support from transient person power, and that power must shift to a sustained team of people with permanent positions whose primary focus is on these issues. Maintaining the highest quality data and delivering it on rapid timescales that CE needs requires a diverse cohesive team whose knowledge base covers applied physics, engineering, and data science as well as observational astrophysics. Transient teams are vulnerable to institutional memory loss, lack of integration with detector design and development, and the inability to create extensible and automated systems upon which the detectors continual upgrades depend. In terms of cost planning, it must be considered equal to, not mutually exclusive to, nor redundant with, those teams covering hardware development, control design, and noise reduction.

\textbf{Data Access Models:} Searches and extensive parameter estimation have already been performed by several groups independent of the LIGO Scientific Collaboration. These groups have demonstrated that rapid turn-around for key science is possible. \textbf{The Cosmic Explorer concept proposes a very open model for data access, based on the way that the Rubin Telescope and NASA missions are proceeding.} There was discussion that the U.S. approach to Open Data is different from the European approach, especially since the best science comes from Cosmic Explorer operating jointly with Einstein Telescope. \textbf{It was identified that discussion is needed between the U.S. and E.U. partners on the path forward for data access.} A recurring theme was the incentive for experimental scientists to invest effort in 3G if data is open. Is the current LSC model where experimenters are generally detached from the data analysis, but have authorship on the publications sustainable? Being one author in O(1000) might be intellectually rewarding for the first detections, but will that continue for the next decade? Would an open model make it easier for instrument scientists to participate in small teams pursuing analysis projects and allow them to be more connected to observational science? What would the “Cosmic Explorer Lab” look like in an era of Open Data?
\subsection{LIGO Laboratory Perspective}

\textit{Presenter: David Reitze}

Perspectives on current LIGO Laboratory involvement in Cosmic Explorer as well as exploring future paths for further participation and integration with CE were presented.  The LIGO Laboratory effectively functions as the US national laboratory for gravitational-wave detector development and observatory operations, and is thus poised to play a foundational role in CE as it moves forward.  LIGO Laboratory staff are already playing key roles in many CE activities, including the overall leadership of the CE Horizon Study Team and Consortium as well as R\&D programs for the CE vacuum system.  

The LIGO Lab brings unique expertise in areas critical to the success of CE, including  interferometric gravitational-wave detector R\&D and the management of large-scale observatory construction projects, having successfully built the LIGO Hanford and Livingston Observatories and the Initial and Advanced LIGO detectors.  The scale and scope of CE will require much larger levels of participation and effort than the LIGO Laboratory can provide, however its capabilities in key areas will be essential for CE to move forward on the timescales envisioned in the Horizon Study.   \textbf{To foster the engagement of the LIGO Lab in CE, discussions should begin soon among the CE team, the LIGO Laboratory management team, and the NSF.}

\medskip
\noindent Other points covered in the presentation included:
\smallskip

\textit{International contributions to CE} - the Advanced LIGO Project benefited greatly from in-kind hardware contributions from institutions in Europe and Australia.  Given CE’s much larger scale and its estimated cost, partnerships with international institutions to lead the R\&D on key CE subsystems and deliver those subsystems will leverage CE development and construction costs as well as demonstrate the commitment of the international community toward CE as a pillar of the next generation ground-based gravitational-wave observatory network.  Discussions should begin soon to identify the  scope and scale of international contributions.  

\textit{CE construction funding} - \textbf{a major challenge going forward is the need to greatly increase the NSF Major Research Equipment and Facilities Construction (MREFC) funding line needed to support CE construction.}While CE presents an extremely compelling science case, the projected price tag of US\$1.6B ensures that it will be competing with several other large-scale astronomical observatories (TMT, GMT, ngVLA)  and physics experiments (CMB-S4 and IceCube Gen2) for construction funding. Upgrades to the current LIGO 4km detectors (incremental or e.g., for a Voyager-type cryogenic step) would also compete for funds. Support of all of these projects (should they go forward) will necessitate a 3X to 4X increase in the annual MREFC budget over current levels. This will be a `heavy lift’, requiring sustained engagement not only with NSF but also with the US Congress who are responsible for authorizing and appropriating NSF’s budget.  Ultimately, it may very well be the availability of funding that limits the pace of CE. 
\subsection{LIGO Scientific Collaboration perspective}

\textit{Presenter: Patrick Brady}

The LIGO Scientific Collaboration remains focused on making the most effective
use of the current observatories in Hanford, WA and Livingston, LA to observe
the gravitational-wave sky. Over the past 5 years, the field has gone from the
first direct detection of gravitational waves from the merger of a pair of
black holes to a routine detection rate of once per 5 days. The most recent
data release identifies 90 signals likely to be from merging compact objects
including black holes and neutron stars. Other milestones include the first
detection of gravitational waves from a merging neutron star GW\,170817, which
was also the first multi-messenger event observed in gravitational waves, the
introduction of public alerts at the beginning of the third LIGO-Virgo-KAGRA
(LVK) observing run (O3), and an early warning test at the end of O3. The LVK
has also been releasing strain data in chunks of 6 months about 18 months after
the data are acquired. The public alerts are received with great interest by
the broader astronomy community. Other gravitational-wave scientists use these
data to search for additional sources of gravitational waves. 

The success of the current generation of gravitational-wave detectors is
tightly bound with network operations to provide increased detection rates,
improved localization, and access to additional information in the
gravitational waves. LIGO-Virgo-KAGRA are discussing a new organizational
framework for collaborating on observational and instrumental science, for
jointly planning and executing observing runs, and to share common services.
The charter and governance structure of this organization, called the
International Gravitational-Wave Observatory Network (IGWN), are being
developed. 

Over the next decade, the LSC plans to carry out a sequence of observing runs
with improving sensitivity. The fourth (O4) observing run will begin in
December 2022 with a binary neutron star range of 160-190 Mpc. The fifth (O5)
observing run will follow around the middle of the decade with a planned
sensitivity more than 300 Mpc. Evolution in the theoretical waveforms, the approaches to `cleaning' the data, the analysis pipelines, and the parameter estimation will be pursued in parallel with the instrument advances.
\textbf{These improvements through the O5 LVK Observing run will bring the detection rate
of compact binaries to several per day as we approach the end of the decade and
may reveal other sources of gravitational waves, such as those from pulsars or
a stochastic background of gravitational waves.} 

The LSC and the LIGO Lab are currently examining options for additional instrumental upgrades to be implemented after O5 that could be installed within the current
LIGO facilities. The goal is a plan that dovetails with Cosmic Explorer
(and Einstein Telescope) construction and initial observations. The research,
development and implementation of upgrades in LIGO (Virgo and KAGRA) can be used
to demonstrate new technologies and mitigate risk for the next generation
facilities. The LSC is committed to continued observational coverage of the
gravitational-wave sky for multi-messenger astrophysics and looks forward to
detecting as yet unobserved gravitational-wave sources. 

\subsection{National Science Foundation perspective}

\textit{Presenter: Pedro Marronetti; Summarized by Shoemaker}
\bigskip

The US National Science Foundation has been the source for all US funding of NSF's LIGO and for the greater ground-based gravitational-wave program in the US. Its support continued beyond the initial construction of the Livingston and Hanford 4km Observatories, through the major Advanced LIGO upgrade, and is now enabling the A+ incremental upgrade. Very welcome additional support has come from the UK, Germany, and Australia, and the worldwide R\&D effort has been crucial in developing the concepts which the Caltech/MIT LIGO Laboratory has engineered, installed, commissioned, and operated. 

The NSF funding mechanism for large projects is the Major Research Equipment and Facilities Construction (MREFC) account, currently targeting projects of US\$70 million and greater. Initial planning and design and post-construction operation and maintenance are supported through the Research and Related Activities (R\&RA) account. Current MREFC projects include ALMA and the Vera Rubin Observatory. 

There is a well-defined process for developing a project to the point of readiness for MREFC funding, and subsequently through the Conceptual, Preliminary, and Final design phases. The process is described in some detail in the MFG.\footnote{\url{https://www.nsf.gov/bfa/lfo/lfo_documents.jsp}} \textbf{To note is that the NSF follows a policy of no additional funding after the initial MREFC allocation; the project must be de-scoped in case of projection of overruns.} Broader impacts are considered as important core elements of Projects, and are funded as part of the MREFC. 

A rough likely timeline was given for Cosmic Explorer. With the completion of the CE Horizon Study, the NSF can now look for community support and `blue ribbon panel' support, which could enable the NSF Director to authorize funding for the Conceptual Design (CD) phase. The CD and Preliminary Design phases are normally 2-3 years each; if successful, the NSF could recommend to Congress to fund the MREFC and the 2-3 year Final Design would be launched. \textbf{The Cosmic Explorer project could take from entrance into the Design Stage to beginning of operations from 11 to 16 years.} Evidently, some steps could go more quickly or more slowly. 

An ongoing challenge for the NSF is managing the operations costs, in particular in the Astronomy Division. LIGO is a significant operations cost for the Physics division. The NSF may in the coming years adopt a different approach, but in any event the accumulated cost of some decades of operation of Cosmic Explorer is a non-negligible consideration for NSF adoption of CE. 

While additional sources of funds (as an example the US Department of Energy) for CE would ease the burden on the NSF and could raise the likelihood and/or speed of execution of the Project, multi-agency Partnerships are complicated to establish and lead to reporting burdens which cover all the engaged funding agencies. 

\subsection{International Collaboration}

\textit{Presenter: Rai Weiss; Summarized by Shoemaker}

Previous Dawn meetings have had extended discussions on the subject of the value of collaboration among the groups developing the next generation of gravitational-wave observatories. The advantages are multifold:

\begin{itemize}
\parskip -3pt
    \item Improved efficiency of R\&D due to shared insights, and the possibility to `divide and conquer' by distributing tasks across the world-wide research groups
    \item Developing a common observational science vision, and seeking to develop a network which seeks an optimal use of the resources available worldwide
    \item Growth of the scientific workforce in an environment which stimulates exchanges and intellectual cross-pollination
    \item Leveraging of the investments across countries and continents, ensuring the best use of the significant funds needed to build next-generation observatories
\end{itemize}

However, there are also difficulties. To date, the two leading projects, Einstein Telescope and Cosmic Explorer, have been out of sync, due to the earlier start of Einstein Telescope. ET also was conceived before the first detections and without other GW detectors in mind; CE is designed around the post-initial-detection environment with networking in mind. There are reasonable concerns that the funding for projects will take different timelines and suffer from different detours and a desire not to have all detectors held to the slowest of the timelines. There is also a sense of desire for independence and freedom to pursue the locally preferred solution to any given challenge.

LIGO and Virgo have taken paths which have explored both independence and collaboration in different domains and epochs, and certainly the advantages of collaboration and shared effort are clear.  The data analysis effort is already very well coordinated and fully joint between LIGO and Virgo, and KAGRA is joining that effort. A recent initiative, IGWN (for International Gravitational-Wave Network), focuses to date on the shared LIGO-Virgo-KAGRA computing infrastructure. The distribution of data-analysis computing tasks, efficient handling of low-latency alerts, and management of security and access are profiting from this joint effort. 

The instrument development has been less closely coordinated to date. There is open communication between LIGO, Virgo, and KAGRA, but the differences in infrastructure, timelines, funding styles and constraints, and finally the social styles of LIGO, Virgo, and KAGRA have made it such that the collaboration has been mostly at the level of workshops and meetings rather than substantially shared effort. 

\textbf{It is perfectly clear from the presentations on the observational science the advantages of a true network of detectors with the kind of sensitivity foreseen for the detectors planned for the next generation of observatories. It is crucial to realize that science to have detectors which work as synergistically as is feasible, from the perspectives of frequency range, sensitivity, uptime, and upgrade planning. It is also crucial to make the best possible use of the significant resources -- financial and human -- needed to realize the network. Closer collaboration can help realize these imperatives.} 

Some suggestions for moving adiabatically to a more collaborative environment for ET and CE were made. Starting with common modeling environments, base models for noise sources, and exchanges of input parameters, cross-checks on both the physics of limitations to sensitivity and on design trades will become easy and reduce risk for both CE and ET. From this some common designs could develop, which may be fabricated independently, or ultimately with teams making all of a given component for ET and CE. A near term target that is gaining momentum is joint effort on value engineering of the vacuum systems for ET and CE, with CERN playing a role as well. Close coordination of the run planning should be foreseen, with a group drawn from the ET and CE teams sharing the task and authority. 

Several parallel efforts seem advisable to make progress. Sharing email lists, and giving invitations to meetings, is easy and immediately fruitful. IGWN may provide a structure and a starting point for an organization. GWIC can ask GWAC if there are mechanisms for complementary light-weight funding for meetings, meeting technology, and persons who are funded to facilitate joint effort. 

\textbf{A recommendation is to establish a small group of persons, representing the entities who wish to be involved, who are charged to explore and launch efforts along these lines of initiating coordinated activities between CE and ET.}

%%%%%%%%%%%%%%%%%%%%%%%%%%%%%%%%%%%%%
\section{Realizing Designs}\label{sec:projects}
\subsection{Megaprojects Round Table}

\textit{Gary Sanders, discussion leader/coordinator; 
Participants: Anne Kinney, Natalie Roe, Suzanne Staggs, Sidney Wolff, Jim Yeck}

The megaprojects panel includes experts from other fields of physical science as well as leaders of large projects in astronomy, astrophysics, cosmology, particle and nuclear physics. These experts have led in projects with private support as well as projects sponsored by NSF, NASA, DOE and European public agencies, and national and global in membership. Informed by the very comprehensive Cosmic Explorer Horizon Study and the preceding discussions in the workshop, the panelists provided a number of important perspectives and useful guidance going forward.

The goals for the panel were to suggest a roadmap and the next concrete steps for Cosmic Explorer. Key to these would be the organizational structure for the design phase and its evolution to the construction and operations phase. The current Horizon Study is led by a university group. While it represents an excellent step, the CE project is seen as a \$2 billion USD (2030 \$) project that will require, beyond the financial resources, strong institutional capabilities and robust governance. It was felt that \textbf{the design phase of the project should urgently set an organization in place that represents a step on a path to the ultimate appropriate robust organization.} Financial, institutional and intellectual resources needed suggest models that may involve the LIGO Laboratory, management and operating contractors of the type currently acting for NSF, national laboratories or new corporations. The roadmap should define a clear trajectory to develop, establish and exercise appropriate governance. 

\textbf{A driver of the project will be selection of the site.} This process should be carried out so that it has completed the requirements, solicitation, proposal reviews and basis for selection by the end of the design phase. Demonstration of compliance with environmental and historical preservation regulations as well as  cultural public participation processes should be at the decision stage. However, it is well recognized that true engagement with the local candidate communities will be critical and this must start at the earliest possible time; it does not suffice simply to have legal clearance to proceed.  The project’s presence and face must engage and partner with local communities to facilitate CE as a joint proposal of the original proponents and the host communities. This is a daunting challenge but needed to lay the foundation for a successful CE project. This theme was well understood and discussed by the proponents and the panel alike. It was referred to as “integration” with the host community.

\textbf{The combination of defining the governance and the collaboration with the host communities (we assume 2 communities for 2 sites) defines “who” CE is. This is key.}

The panelists described these processes for prior large projects. Based upon projects like ALMA, SKA, and TMT, this process takes a decade.

Local communities are not monoliths. Host community residents and local indigenous peoples may represent distinct local groups who must be engaged. But these groups are likely not monolithic as well and engagement, opposition, acceptance and consent may be quite varied among these multiple groups. \textbf{The complexity and centrality of interactions between the CE project and the host communities must be addressed by genuine presence and sincere close engagement.}

Though the legal processes such as environmental impact statements and cultural consultations may lead to legal approval of CE, these are not substitutes for the intimate engagement described above. However, the leadership and involvement of the government agencies in these public participation processes has been found to add gravitas and trust to commitments made to host communities.

It was felt that \textbf{the CE design stage should move early to a baseline configuration and fully develop that design rather than continuing the trade studies of potential design options}. A fully developed baseline design, complete with system engineering, scope definition, risk management and reliable cost and schedule studies is a superior point of departure for any design variations that may develop out of necessity, such as budget ceilings from sponsors. 

Given the size of the project, while a lead agency such as NSF may dominate in support, multiple agencies and partners may be needed to accomplish a project of the scale and complexity of CE. Private funding, given its agility, may be very helpful in getting the design stage off to a timely start.

DOE laboratories possess key skills in civil construction and high vacuum technology, cost drivers for CE. DOE laboratories should be engaged but not as service organizations in these critical skills. The science reach of CE covers many key thrusts in the DOE high-energy, nuclear and particle astrophysics realms. The CE community should be out giving talks and continuing their participation in the Snowmass process.

CE will drive NSF to consider the needed budget size, growing NSF capabilities. Community engagement should extend beyond host community integration to include partners such as Historically Black Colleges and Universities and other communities for whom CE may leverage participation these sciences.

LIGO has shown that it can generate a steady influx of the brightest young people. CE can be a catalyst for frontier activity and involvement at universities. University involvement generates diversity as well as novelty and influence in other fields. And a careful look at university engagement in LIGO shows that it created many of the key successes of LIGO.

As NSF is likely the lead sponsor, lessons can be learned beyond LIGO, from the history of the astronomy observatories. Site selection, host community support and opposition including legal challenges, managing organizations with experience and trust with the NSF; these are aspects to be carefully considered.

\textbf{The panel felt that, with these issues in mind, CE should move out promptly.}

\subsection{Recommendations and Conclusions}

\textit{Presenter: Dave Reitze}
\bigskip

A high level overview of the workshop themes was presented to the participants at the Dawn VI close out, followed by a moderated discussion among the participants. Participants felt that the CE science case was strong and well supported in the Horizon Study.  Most of the science goals are robust against discoveries that may occur with the current generation of detectors.  Surveying the ‘heavy’ mass spectrum of black holes (with masses greater than a few hundred solar masses) is one such example. 

Care will be needed to make a clear and compelling case for a standalone CE, while noting the added gains and discovery science that come with partnering with ET. 

Although it is clear that the CE science case is strong, the science case is a necessary but not sufficient condition to enable CE to move forward toward securing funding, being constructed, and into operations.  As noted in Dawn VI presentations and discussions, the  complexity and scale of the project in terms of infrastructure, site selection and acquisition, cost, and needed human resources is such that there will be many paths by which CE could be delayed or, much worse, fail to be built. 

Some recommendations can be enunciated:

\begin{itemize}
\parskip -3pt

\item A very significant upgrade to the current LIGO observatories may well delay the start of funding of CE.
\item It is important to communicate clearly that some key CE science targets rely wholly or partially on GW-enabled multi-messenger astronomy -- thus requiring a three-site GW observatory network for optimal source localization. While CE and ET collectively make up three-site network each science case must ‘stand on its own’ to secure construction funding as there is no guarantee that both will be operational.

\item Drawing lessons learned by studying other projects of similar scales will prove valuable for the CE Project as it scales up.

\item In the near term, there is a strong desire from the broad community that the present detectors continue to observe until CE and ET are available. 
\item GW facilities should give a ``mission statement” level of support (if not an absolute guarantee) to seek to continue operations during the main mission lifetimes of space missions dedicated to GW follow-up. This requires a rough 5-10 year horizon for the planning of instrument operation, to match the time scale for space missions.

\item In the near term, incremental deepening of the collaboration between ET and CE is to the advantage of all parties. Point persons in ET and CE should be named to bring this to fruition.
\end{itemize}

Summarizing the two main conclusions from the meeting: 
\textbf{\begin{itemize}
\parskip -3pt
    \item The science opportunities afforded by CE and ET are broad and compelling, impacting a wide range of disciplines in physics and high energy astrophysics.
    \item A strong endorsement of Cosmic Explorer, as described in the CE Horizon Study, is a primary outcome of DAWN VI.
\end{itemize}}

%\newpage

%%%%%%%%%%%%%%%%%%%%%%%%%%%%%%%%%%%%%%%%
% bibliography
%\bibliographystyle{unsrt}
%\bibliography{references}

\begin{appendices}
\break
\section{Contributors to and endorsers of this report}\label{appendix}

\begin{longtable}{p{.37\textwidth}  p{.63\textwidth} }

Naresh	Adhikari	&	University of Wisconsin Milwaukee	\\
Odylio	Aguiar	&	Instituto Nacional de Pesquisas Espaciais	\\
Ryan	Andersen	&	Univ. California, Riverside	\\
M. Celeste	Artale	&	University of Padua	\\
K. G.	Arun	&	Chennai Mathematical Institute	\\
Stefan W.	Ballmer	&	Syracuse University	\\
Enrico	Barausse	&	SISSA (Trieste, Italy)	\\
Barry	Barish	&	Caltech/LIGO and UCR	\\
Lisa	Barsotti	&	MIT	\\
Matteo	Barsuglia	&	CNRS - APC 	\\
Willke	Benno	&	Leibniz University Hanover	\\
Beverly	Berger	&	Stanford	\\
Edo	Berger	&	Harvard	\\
Emanuele	Berti	&	JHU	\\
GariLynn	Billingsley	&	California Institute of Technology	\\
Ofek	Birnholtz	&	Bar-Ilan University	\\
Sebastien	Biscans	&	Caltech and MIT 	\\
Sylvia	Biscoveanu	&	MIT	\\
Marie Anne	Bizouard	&	CNRS - Observatoire de la Cote d'Azur	\\
Alexey	Bobrick	&	Technion Institute	\\
Sukanta	Bose	&	IUCAA and Washington State University	\\
Patrick	Brady	&	University of Wisconsin-Milwaukee	\\
Marica	Branchesi	&	Gran Sasso Science Institute	\\
Jim	Brau	&	University of Oregon	\\
Floor	Broekgaarden	&	Center for Astrophysics - Harvard \& Smithsonian	\\
Duncan 	Brown	&	Syracuse University	\\
Daniel	Brown	&	University of Adelaide	\\
Tomasz	Bulik	&	University of Warsaw	\\
Alessandra	Buonanno	&	Max Planck Institute for Gravitational Physics, Potsdam	\\
Laura 	Cadonati	&	Georgia Tech	\\
Elenna	Capote	&	Syracuse University	\\
Marco	Cavaglia	&	Missouri S\&T	\\
Poonam	Chandra	&	National Centre for Radio Astrophysics, TIFR, India	\\
Hsin-Yu	Chen	&	Massachusetts Institute of Technology	\\
Xu	Chen	&	University of Western Australia	\\
Giacomo	Ciani	&	University of Padova	\\
Alessandra	Corsi	&	Texas Tech University	\\
Michael	Coughlin	&	UMN	\\
David	Coulter	&	University of California, Santa Cruz	\\
Jolien	Creighton	&	University of Wisconsin–Milwaukee	\\
Alexander	Criswell	&	University of Minnesota Twin Cities	\\
Alan	Cumming	&	University of Glasgow, UK	\\
Kristen	Dage	&	McGill University	\\
Stefan 	Danilishin	&	Maastricht University/Nikhef, the Netherlands	\\
Saurya	Das	&	University of Lethbridge	\\
Derek	Davis	&	Caltech	\\
Nicholas	Demos	&	MIT	\\
Riccardo	DeSalvo	&	Ric-lab LLC	\\
Mariano	Dominguez	&	Universidad Nacional de Cordoba	\\
Daniela	Doneva	&	University of Tuebingen	\\
Bruce	Edelman	&	University of Oregon	\\
Anamaria	Effler	&	Caltech, Pasadena, CA	\\
Matthew	Evans	&	MIT	\\
Jose María	Ezquiaga	&	University of Chicago	\\
Stephen	Fairhurst	&	Cardiff University	\\
Ke	Fang	&	University of Wisconsin-Milwaukee	\\
Maxime	Fays	&	ULi\`ege	\\
Svenja	Fleischer	&	Western Washington University	\\
Wen-fai 	Fong	&	CIERA, Northwestern University	\\
Giacomo	Fragione	&	Northwestern University	\\
Raymond	Frey	&	University of Oregon	\\
Josh	Frieman	&	Fermilab	\\
Peter	Fritschel	&	M.I.T.	\\
Chris	Fryer	&	LANL	\\
Hannah	Gallagher	&	RIT	\\
Gianluca	Gemme	&	INFN Sezione di Genova	\\
Andrew	Gendre	&	OzGrav-UWA	\\
Andrew	Geraci	&	Northwestern University	\\
Priyanka 	Giri	&	INFN Pisa	\\
Amy	Gleckl	&	California State University Fullerton	\\
Gabriela	Gonzalez	&	Louisiana State University	\\
Aniello	Grado	&	INAF	\\
Alexandra	Gruson	&	California State University Fullerton	\\
Anuradha	Gupta	&	University of Mississippi	\\
Evan	Hall	&	Massachusetts Institute of Technology	\\
Sophia	Han	&	UC Berkeley/INT, U Washington	\\
Mark	Hannam	&	Cardiff University	\\
Jan	Harms	&	Gran Sasso Science Institute	\\
Gregory	Harry	&	American University	\\
Carl-Johan	Haster	&	Massachusetts Institute of Technology	\\
Jack	Heinzel	&	MIT	\\
Jackie	Hewitt	&	MIT	\\
Daniel	Holz	&	UChicago	\\
Chuck 	Horowitz	&	Indiana University	\\
Jim	Hough	&	University of Glasgow, UK	\\
Panagiotis	Iosif	&	Aristotle University of Thessaloniki	\\
Bala	Iyer	&	ICTS-TIFR, Bangalore	\\
Wenxuan	Jia	&	MIT	\\
Nathan	Johnson-McDaniel	&	University of Mississippi	\\
Li	Ju	&	University of Western Australia	\\
Vicky	Kalogera	&	Northwestern University	\\
Jonah	Kanner	&	Caltech	\\
Mansi	Kasliwal	&	Caltech	\\
Stavros	Katsanevas	&	European Gravitational Observatory	\\
Erik	Katsavounidis	&	Massachusetts Institute of Technology	\\
Keita	Kawabe	&	LIGO Hanford Observatory	\\
Joey Shapiro	Key	&	University of Washington Bothell	\\
Jeffrey	Kissel	&	LIGO Hanford Observatory	\\
Kostas	Kokkotas	&	University of Tuebingen	\\
Antonios	Kontos	&	Bard College	\\
Mikhail	Korobko	&	University of Hamburg, Germany	\\
Kevin	Kuns	&	MIT	\\
Arjun	Kurur	&	Indian Institute of Technology, Madras	\\
Harald	L\"uck	&	Leibniz Universit\"at Hannover	\\
Adele	La Rana	&	INFN - Rome 1 \& University of Verona	\\
Philippe	Landry	&	Canadian Institute for Theoretical Astrophysics 	\\
Michael	Landry	&	Caltech - LIGO Hanford Observatory	\\
Brian	Lantz	&	Stanford University	\\
Paul	Lasky	&	Monash University	\\
Albert	Lazzarini	&	California Institute of Technology	\\
Hyung Won	Lee	&	Inje University, Korea	\\
Hyung Mok	Lee	&	Seoul National University	\\
Luis	Lehner	&	Perimeter Institute	\\
Giovanni	Losurdo	&	INFN - Pisa	\\
Hudson	Loughlin	&	MIT	\\
Geoffrey	Lovelace	&	Cal State Fullerton	\\
Ronaldas	Macas	&	University of Portsmouth	\\
Lorena	Maga\~na Zertuche	&	University of Mississippi	\\
Michele 	Maggiore	&	Universit\'e de Gen\`eve	\\
Vuk	Mandic	&	University of Minnesota Twin Cities	\\
Szabolcs	Marka	&	Columbia University in the City of New York	\\
Zsuzsa	Marka	&	Columbia Astrophysics Laboratory	\\
Rodica	Martin	&	Montclair State University	\\
Daniel	Martinez	&	CSU fulelrton	\\
Kenneth 	Mason	&	Massachusetts Institute of Technology	\\
Nergis	Mavalvala	&	Massachusetts Institute of Technology	\\
David	McClelland	&	The Australian National University	\\
Lee	McCuller	&	California Institute of Technology	\\
Kara 	Merfeld 	&	University of Oregon	\\
Yuta	Michimura	&	University of Tokyo	\\
Cole	Miller	&	UMd	\\
Edoardo	Milotti	&	University of Trieste and INFN-Sezione di Trieste - Italy	\\
Guenakh	Mitselmakher	&	University of Florida	\\
Richard	Mittleman	&	MIT	\\
Marlo	Morales	&	California State University, Fullerton	\\
Ewald	Mueller	&	Max-Planck-Institute for Astrophysics	\\
Suvodip	Mukherjee	&	Perimeter Institute	\\
Arunava	Mukherjee	&	Saha Institute of Nuclear Physics	\\
Rohit	Nair	&	Model College 	\\
Lan Quynh	Nguyen	&	University of Notre Dame	\\
Alexander	Nitz	&	Max Planck Institute for Gravitational Physics, Hannover	\\
Benjamin	Owen	&	Texas Tech University	\\
Archana	Pai	&	Indian Institute of Technology Bombay archanap@iitb.ac.in	\\
Oli	Patane	&	California State University Fullerton	\\
Hiranya	Peiris	&	University College London / Stockholm University	\\
M.E.S	Pereira	&	Universität Hamburg	\\
Harald	Pfeiffer	&	Max Planck Institute for Gravitational Physics, Potsdam	\\
Jorge	Piekarewicz	&	Florida State University	\\
Michele 	Punturo	&	INFN Perugia	\\
Frederick	Raab	&	California Institute of Technology	\\
Jocelyn	Read	&	California State University, Fullerton	\\
Stuart	Reid	&	University of Strathclyde	\\
Dave 	Reitze	&	Caltech	\\
Michael	Rezac	&	Cal State-Fullerton	\\
Fulvio	Ricci	&	INFN Sezione di Roma	\\
Jonathan	Richardson	&	University of California, Riverside	\\
Keith	Riles	&	University of Michigan	\\
Jameson	Rollins	&	LIGO Caltech	\\
Michael	Ross	&	University of Washington	\\
Surabhi	Sachdev	&	UWM	\\
Mairi	Sakellariadou	&	King's College London	\\
Gary H. 	Sanders	&	Simons Observatory, Center for Astrophysics and Space Sciences, UC San Diego	\\
Misao	Sasaki	&	University of Tokyo	\\
Bangalore 	Sathyaprakash	&	Penn State and Cardiff University	\\
Richard	Savage	&	LIGO Hanford Observatory	\\
Robert	Schofield	&	University of Oregon	\\
Audrey P.	Scott	&	University of Chicago	\\
Olga	Sergijenko	&	Taras Shevchenko National University of Kyiv \& MAO NASU	\\
Lijing	Shao	&	Peking University	\\
Peter	Shawhan	&	University of Maryland	\\
Deirdre	Shoemaker	&	University of Texas at Austin	\\
David	Shoemaker	&	MIT	\\
Vasileios	Skliris	&	Cardiff University	\\
Joshua	Smith	&	California State University, Fullerton	\\
Frank	Soboczenski	&	King's College London	\\
Varun	Srivastava	&	Syracuse University	\\
Andrew 	Steiner	&	University of Tennessee, Knoxville \& Oak Ridge National Laboratory	\\
Sebastian	Steinlechner	&	Maastricht University, the Netherlands	\\
Riccardo	Sturani	&	International Institute of Physics, Natal (Brazil)	\\
Shreevathsa	Chalathadka Subrahmanya	&	University of Hamburg	\\
Ling	Sun	&	The Australian National University	\\
Patrick	Sutton	&	Cardiff University	\\
Satoshi	Tanioka	&	Syracuse University	\\
David	Tanner	&	University of Florida	\\
Christina	Th\"one	&	IAA - CSIC	\\
Eleonora	Troja	&	University of Rome Tor Vergata	\\
Rhondale	Tso	&	Caltech	\\
Elias C.	Vagenas	&	Kuwait University	\\
Gabriele	Vajente	&	Caltech	\\
Jo	van den Brand	&	Nikhef and Maastricht University	\\
Peter	Veitch	&	OzGrav-Adelaide	\\
Subham	Vidyant	&	Syracuse University	\\
Salvatore	Vitale	&	MIT	\\
Alan	Weinstein	&	California Institute of Technology	\\
Chris	Whittle	&	Massachusetts Institute of Technology	\\
Helvi	Witek	&	University of Illinois Urbana-Champaign	\\
Victoria	Xu	&	MIT	\\
Huan	Yang	&	Guelph University	\\
Stoytcho	Yazadjiev	&	University of Sofia	\\
Nicolas	Yunes	&	UIUC	\\
Liyuan	Zhang	&	California Institute of Technology	\\
Aaron	Zimmerman	&	University of Texas at Austin	\\
Michael	Zucker	&	Caltech and MIT 	\\

\end{longtable}
 
\end{appendices}

\end{document}